\documentclass[prc, amsfonts, amssymb, amsmath, preprintnumbers, twocolumn, showkeys, nofootinbib,superscriptaddress,floatfix]{revtex4-1}%
\usepackage[english]{babel}
\usepackage[colorinlistoftodos, color=green!40, prependcaption]{todonotes}
\usepackage{amsmath}
\usepackage{amssymb}
\usepackage{booktabs} %
\usepackage{hyperref}

\usepackage[compat=1.1.0]{tikz-feynman}
\usepackage{contour}
\usepackage{subfigure}  

\usepackage{amsthm}
\usepackage{mathtools}
\usepackage{physics}
\usepackage{xcolor}
\usepackage{graphicx}
\usepackage[left=23mm,right=13mm,top=35mm,columnsep=15pt]{geometry} 
\usepackage{adjustbox}
\usepackage{placeins}
\usepackage[T1]{fontenc}
\usepackage{lipsum}
\usepackage{csquotes}

\newcommand\pT{\ensuremath{p_\mathrm{T}}}

\newcommand\pTjet{\ensuremath{p_\mathrm{T}^{jet}}}

\newcommand\met{\ensuremath{E_{T}^{\mathrm{miss}}}}
\begin{document}
\title{Charm jets as a probe for strangeness at the future Electron-Ion Collider}

\author{Miguel Arratia}
\affiliation{Department of Physics and Astronomy, University of California, Riverside, CA 92521, USA}
\affiliation{Thomas Jefferson National Accelerator Facility, Newport News, VA 23606, USA}
\author{Yulia Furletova}
\affiliation{Thomas Jefferson National Accelerator Facility, Newport News, VA 23606, USA}
\author{T.~J.~Hobbs}
\affiliation{Department of Physics, Southern Methodist University, Dallas, TX 75275, USA}
\affiliation{Jefferson Lab, EIC Center, Newport News, VA 23606, USA}
\author{Fredrick  Olness}
\affiliation{Department of Physics, Southern Methodist University, Dallas, TX 75275, USA}
\author{Stephen J. Sekula}
\email{corresponding author: ssekula@smu.edu}
\affiliation{Department of Physics, Southern Methodist University, Dallas, TX 75275, USA}

\date{\today} %

\begin{abstract}
We explore the feasibility of the measurement of charm-jet cross sections in charged-current deep-inelastic scattering at the future Electron-Ion Collider. This channel provides clean sensitivity to the strangeness content of the nucleon in the high-$x$ region. 
We estimate charm-jet tagging performance with parametrized detector simulations. We show the expected sensitivity to various scenarios for strange parton distribution functions. We argue that this measurement will be key to future QCD global analyses, so it should inform EIC detector designs and luminosity requirements. 
\end{abstract}

\preprint{JLAB-PHY-20-3205, SMU-HEP-20-05}

\maketitle

\section{Introduction}
\label{sec:intro}

The future Electron-Ion Collider (EIC)~\cite{Accardi:2012qut} will herald a new era for the study of nucleon structure by producing a unique data sample
of deep-inelastic scattering (DIS) measurements off protons, deuterium, and helium with high polarization of both beams.  The EIC will undertake a comprehensive mapping of the nucleon's multidimensional {\it tomography}, with the broad goal
of unlocking the proton's partonic substructure. For collinear quantities such as the unpolarized and nucleon-helicity dependent parton distribution
functions (PDFs), $f(x,Q)$ and $\Delta f(x,Q)$, this entails heightened precision for unraveling flavor and kinematical ({\it i.e.},
$x, Q$) dependence.

The EIC will possess unprecedented capabilities to address these issues, owing mainly to high luminosities (100--1000 times the
HERA instantaneous luminosity). Moreover, given its lower center-of-mass energy, the EIC's coverage will extend to very high $x$
--- reaching a factor of 10 higher $x$ for a specified $Q^{2}$ relative to HERA. As such, the EIC will be well-disposed
to exploring not only the gluon-dominated region at lower $x$, but also the high-$x$ frontier.

This access to high $x$ will allow the EIC to resolve long-standing questions regarding the precise balance of quark flavors
contributing to the proton's valence-region structure. Disentangling high-$x$ flavor dependence in PDFs poses a challenge due to the rapid decline of even the valence quark distributions beyond $x\! \gtrsim\! 0.1$ and the comparatively small normalization of the $d$-quark and sea PDFs relative to the $u$-quark density. Extractions of the $d$-type PDFs are further complicated by nuclear corrections needed for DIS off deuterium. 
From the perspective of non-perturbative dynamics, the flavor
decomposition of the proton (and of other light hadrons) carries signatures of QCD's patterns of symmetry breaking, including dynamical chiral symmetry breaking
and the low-energy violation of flavor-$\mathrm{SU}(3)$ symmetry. 

This latter issue has stimulated considerable effort in separating nucleon strangeness, $s,\bar{s}(x,Q)$,
from the rest of the nucleon's light-quark sea, $\bar{u}, \bar{d}(x,Q)$. The strange PDF is often examined in terms of its fractional size relative to the 
total $\mathrm{SU}(2)$ quark sea,\footnote{%
   For reference, other quantities used may be the fractional size relative to the $d$-quark sea
   $r_s(x,Q)={(s+\bar{s})/(2\bar{d})}$, or the integrated momentum  ratio 
   $\kappa(Q)={[\int x (s+\bar{s}) dx]} / {[\int x (\bar{u}+\bar{d}) dx]} $.}
$R_s$, 
\begin{equation}
R_s(x,Q) = \frac{s(x,Q) + \bar{s}(x,Q)}{\bar{u}(x,Q) + \bar{d}(x,Q)}\ .
\label{eq:Rs}
\end{equation}
A primary source of information in contemporary determinations of the strange PDF is supplied by fixed-target neutrino DIS experiments involving
heavy nuclear targets. The interpretation of data from these experiments is complicated by a subtle interplay of effects arising from nuclear, target-mass,
and other power-suppressed corrections, as well as potential contamination from target fragmentation~\cite{Schienbein:2009kk,Kovarik:2010uv}. 
These effects present a serious challenge to rigorously quantifying the uncertainty of the subsequent PDF extraction.

Sensitivity to the strange PDF can also be gained with measurements of identified hadrons in the semi-inclusive DIS (SIDIS) approach~\cite{Aschenauer:2019kzf}. This method has
multiple associated challenges as well, including a strong dependence of the extracted PDF upon the associated fragmentation function or hadronization model.
This issue has prompted efforts to perform simultaneous determinations of PDFs and fragmentation functions~\cite{Sato:2019yez}. A feature of these studies
is that the entanglement of the non-perturbative PDF and fragmentation in a single measurement leads to strong correlations between them.

We note that dimuon measurements in neutrino DIS experiments~\cite{Bazarko:1994tt,Goncharov:2001qe,Olness:2003wz,Samoylov:2013xoa,KayisTopaksu:2011mx} typically
prefer a low value of $R_s$ in Eq.~(\ref{eq:Rs}), whereas kaon SIDIS measurements prefer an even lower value~\cite{Airapetian:2008qf,Alekseev:2009ac}.
In contrast, recent electroweak boson measurements at the LHC prefer a value of $R_{s}$ consistent with unity~\cite{Aad:2012sb,Chatrchyan:2013uja,Aad:2014xca}. New data
are required to understand the apparent tension between these measurements. We discuss the preferences of different contemporary data sets on the unpolarized
strangeness in further detail in Sec.~\ref{sec:Rs}.

In the spin-polarized sector, knowledge of the strange helicity distribution --- a quantity even less constrained than the unpolarized strange PDF --- is crucial to elucidating the origin of the nucleon spin, which remains an unresolved problem~\cite{Deur:2018roz}, which the EIC will elucidate~\cite{Aschenauer:2015ata}. 
In addition to fundamental knowledge of nucleon structure, the strangeness content could also illuminate the dynamics of core-collapse supernova explosions by constraining neutrino-nucleon elastic cross sections~\cite{Hobbs:2016xlg}.  
Studies of strange helicity~\cite{deFlorian:2009vb} have relied on kaon measurements in SIDIS measurements from HERMES~\cite{Airapetian:2008qf} and COMPASS~\cite{Alekseev:2009ac}. Like the
corresponding analyses in the unpolarized sector, these measurements were prone to biases and ambiguities from the needed input from fragmentation functions.

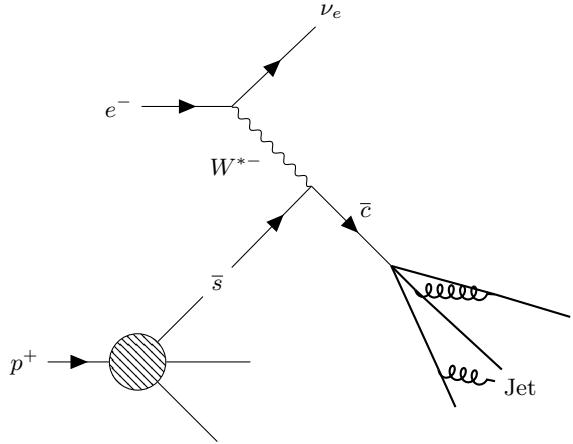
\begin{figure}
    \centering
       \scalebox{1.0}{
    \begin{tikzpicture}[scale=0.5]
     \begin{feynman}
      \vertex(a) {\(e^{-}\)};
      \vertex[right=of a] (b);
      \vertex[above right=of b] (f1){\(\nu_{e}\)};
      \vertex[below right=of b] (c);
      \vertex[below left=of c] (strange){\(\overline s\)};
      \vertex[below right=of c] (charm);
      \vertex[blob, below left=of strange] (dis) {\contour{gray}{}};
      \vertex[left=of dis] (p) {\(p^+\)};
      \vertex[right=of dis] (frag1);
      \vertex[below right=of dis] (frag2);
      \vertex (jet1a) at (10, -5);
      \vertex (jet1b) at (12, -5.6);
      \vertex[below right=0.45cm of charm] (jet2a);
      \vertex[below right=of jet2a] (jet2b) {Jet};
      \vertex (jet3a) at (8.5, -7);
      \vertex (jet3b) at (8.95, -8);
      \diagram* {(a)-- [fermion] (b)-- [fermion] (f1),
      (b)-- [boson,edge label'=\(W^{*-}\)] (c),
      (c)-- [anti fermion] (strange),
      (c)-- [fermion, edge label=$\overline c$] (charm),
      (p)-- [fermion] (dis) -- {(strange), (frag1), (frag2)},
      (charm) -- [thick] {(jet1a),(jet2a), (jet3a)},
      (jet1a) -- [thick] (jet1b),
      (jet2a) -- [thick, gluon] (jet1a),
      (jet2a) -- [thick](jet2b),
      (jet3a) -- [thick] (jet3b),
      (jet3a) -- [gluon,thick] (jet2b),
      };
     \end{feynman}
    \end{tikzpicture}
   }

\caption{\label{fig:feynman_diagram} Leading-order diagram for the production
of final-state charm in charged-current electron-proton DIS.
}
\end{figure}

As we shall demonstrate in this analysis, due to the availability of channels in which the internal structure of the free proton is directly probed by electroweak currents, the EIC has the potential to avoid many of the complications described above. EIC measurements thus represent a unique opportunity to open a
new era of sensitivity to intrinsic strangeness in the proton.
In particular, an alternative way to achieve flavor sensitivity {\it without} fragmentation functions and nuclear corrections is charged-current (CC) DIS
(Fig.~\ref{fig:feynman_diagram}).

Inclusive charged-current and neutral-current (NC) DIS at HERA have been instrumental in proton structure studies~\cite{Abramowicz:2015mha}.
Single jet production in inclusive CC DIS was measured by the ZEUS collaboration~\cite{Chekanov:2003jd,Chekanov:2008af}. Gehrman {\it et al}.~described the ZEUS data
with N$^{3}$LO calculations~\cite{Gehrmann:2018odt}; this work showed that the inclusion of higher-order pQCD corrections stabilized scale variations to the
(sub)percent-level. More recently, the ZEUS collaboration published the first measurement of charm-tagged events in CC DIS~\cite{Abt:2019ngj}. While limited
in precision, the ZEUS work demonstrated that this channel offers a more direct way to access the strange PDF. 

Previous feasibility studies of jet measurements at the EIC focused on NC DIS~\cite{Arratia:2020nxw,Arratia:2019vju,Page:2019gbf,Zheng:2018ssm} and photo-production~\cite{Aschenauer:2019uex,Chu:2017mnm}. The feasibility of inclusive CC DIS measurements at the EIC has been studied by Aschenauer et al.~\cite{Aschenauer:2013iia}. Here, we specifically focus on charm-jet production in CC DIS and its
potential sensitivity to the strange quark sea. This work also differs from recent work by Abdolmaleki et al~\cite{Abdolmaleki:2019acd}, which emphasized the low-$x$ region that would be provided by the Large Hadron-Electron Collider (LHeC)~\cite{AbelleiraFernandez:2012cc}. In contrast, we focus on the valence region, which remains poorly constrained.

The remainder of this article is structured as follows. In Sec.~\ref{sec:fastsim}, we describe the details of the \textsc{Pythia8} and \textsc{Delphes} simulations upon which this
analysis is based. We then (Sec.~\ref{sec:tagging}) discuss the specifics of the charm-jet tagging essential to identifying final-state charm in our CC DIS simulations and then
(Sec.~\ref{sec:strange}) describe the strange-sea PDF inputs we employ to test the event-level discriminating power of charm-jet measurements. Finally, in Sec.~\ref{sec:detreq},
we outline specific detector recommendations and requirements to optimize the sensitivity of CC DIS charm-jet production before concluding in Sec.~\ref{sec:conclusions}.

\section{Simulation}
\label{sec:fastsim}

 We use \textsc{Pythia8}~\cite{Sjostrand:2007gs} to generate 20 million CC DIS events in unpolarized electron-proton collisions with beam energies of 10~GeV and 275~GeV respectively, which is the configuration that maximizes the luminosity in the nominal eRHIC design~\cite{EICdesign}. We enable particle decay anywhere within the proposed tracking volume of an EIC baseline detector, encompassing a cylinder of radius 80 cm and $z$ half-length of 100 cm. We do not include QED radiative corrections, which are relevant in some kinematic regions~\cite{Aschenauer:2013iia}, as these do not drastically affect the focus of our work: the projected precision for strangeness measurements. Moreover, proper treatment of QED radiative corrections requires detailed simulations of detector response that are outside the scope of this work. 
 
In addition, higher-order QCD effects are an important consideration for CC DIS charm production,
which receives corrections at NLO from boson-fusion channels unavailable at LO. Still, we expect these higher-order
corrections to not dramatically alter the kinematic properties used here for the reconstruction and
sensitivity evaluations.
The total $t$-channel CC DIS cross section for \linebreak
$Q^2\!>\!100\mathrm{~GeV^2}$ is reported by \textsc{Pythia8} to be 14.8 pb, which is similar to the NLO calculation in Ref.~\cite{Aschenauer:2013iia}. 
Comparing the LO vs.\ NLO calculations for the charged-current charm-production
structure functions ({\it e.g.}, $F^{W^-}_{2c}$) indicates that the NLO corrections
are generally relatively mild, especially in the large-$x$ region of relevance to the
measurements discussed in this study \cite{Aivazis:1990pe,Thorne:2000zd}.
This is similar to the situation for Monte Carlo-generated $x$, $Q$, $\eta$, and $\pT$ distributions for NC DIS
charm production, which suggest no substantive differences in the shapes of these distributions between LO and NLO
accuracy~\cite{Harris:1997zq}.
Thus, NLO corrections should not significantly impact the present analysis, and we reserve a more
detailed examination of the higher-order correction effects to future work.

\subsection{Detector response parametrization}

 The basic requirements for a EIC detector have been established in order to explore the impact of possible detector choices on
 the realization of physics goals~\cite{EICHandbook}. The baseline EIC detector consists of an inner charged particle tracking system, an electromagnetic calorimeter, a particle-identification (PID) system, and an hadronic calorimeter. The PID system is envisioned to yield at least $3\sigma$ separation of $\pi^{\pm}$, $K^{\pm}$, and $p^{\pm}$ for momenta between $1\!-\!50\mathrm{~GeV}$, depending on $\eta$. Electron identification will primarily be achieved using the electromagnetic calorimeter. A dedicated muon system has not been excluded but is not specified in the baseline.
 
We use the \textsc{Delphes} framework~\cite{deFavereau:2013fsa} to obtain a parametrized simulation of detector response. We show in Table~\ref{tab:resolutions} the parametrization of momentum, energy, and impact parameter resolution used as input for \textsc{Delphes}.  All simulated systems provide full azimuthal coverage. The inner tracker is immersed in a 1.5 T solenoidal magnetic field.
 
 The tracking efficiency at the EIC is expected to be close to unity with negligible fake rate, given the low event multiplicity and the proposed use of redundant low-mass silicon pixel detectors~\cite{EICHandbook}. We incorporate a conservative estimate of tracking inefficiency of 1--5\% depending on the $\eta$ region, which is also shown in Table~\ref{tab:resolutions}. 

\begin{table}
\begin{ruledtabular}
  \begin{tabular}{lc}
    \multicolumn{2}{c}{Tracking resolution} \\
  $[-1.0, 1.0]$  & 0.5\% $\oplus$ 0.05\%$\times p$  \\       
  $1.0<|\eta|<2.5$ & 1.0\% $\oplus$ 0.05\%$\times p$  \\ 
  $2.5<|\eta|<3.5$ & 2.0\% $\oplus$ 0.01\%$\times p$ \\
  \hline
    \multicolumn{2}{c}{Track Impact Parameter Resolution} \\
    Parameter & Resolution [$\mathrm{\mu m}$] \\
    $d_0$ & 20 \\
    $z_0$ & 20 \\
        \hline
      \multicolumn{2}{c}{Charged Particle Tracking Efficiency [\%]} \\
    $\eta$ &  $\pT=[0.1,1.0]\mathrm{~GeV}$ \hspace*{5mm} $\pT > 1.0 \mathrm{~GeV}$ \\
    $[-3.5, -2.5]$ & 95 \hspace*{20mm} 97  \\
    $[-2.5, -1.5]$ & 96 \hspace*{20mm} 98  \\
    $[-1.5, 1.5]$ & 97 \hspace*{20mm} 99   \\
    $[1.5, 2.5]$ & 96 \hspace*{20mm} 98  \\
    $[2.5, 3.5]$ & 95 \hspace*{20mm} 97  \\
  \hline
    \multicolumn{2}{c}{Electromagnetic Calorimeter} \\
    \multicolumn{2}{c}{($E>0.2\mathrm{~GeV}$)} \\
    $\eta$ & Resolution [\%]\\
    $[-4.0, -2.0]$  &  $\sqrt{E} \times (2.0)  \oplus E \times (1.0)$ \\
    $[-2.0, -1.0]$  &  $\sqrt{E} \times (7.0)  \oplus E \times (1.0)$ \\
    $[-1.0,  1.0]$  &  $\sqrt{E} \times (10.0) \oplus E \times (1.0)$ \\
    $[ 1.0,  4.0]$  &  $\sqrt{E} \times (12.0) \oplus E \times (2.0)$ \\
  \hline
    \multicolumn{2}{c}{Hadronic Calorimeter} \\
    \multicolumn{2}{c}{($E>0.4\mathrm{~GeV}$)} \\
    $\eta$ & Resolution [\%]\\
     $[-4.0, -1.0]$ &  $\sqrt{E} \times (50.0) \oplus E \times (10.0)$ \\
     $[-1.0, 1.0]$  & $\sqrt{E} \times (100.0)  \oplus  E \times (10.0)$\\
     $[1.0, 4.0]$  & $ \sqrt{E} \times (50.0)  \oplus   E \times (10.0)$ \\
   \hline
    \multicolumn{2}{c}{PID performance} \\    
     $K^{\pm}, \, \pi^{\pm}$ & $\ge 3\sigma$ separation in the range  \\
     $[-4.0, -1.0]$ & up to 10 GeV \\ 
     $[-1.0, 1.0]$ & up to 6 GeV \\
    $[1.0, 4.0]$ & up to 50 GeV \\
     $e^{\pm}, \, \pi^{\pm}$ & $\ge 2.4\sigma$ separation (rejection factor 50)  \\
     $\mu^{\pm}, \, \pi^{\pm}$ & $\ge 2\sigma$ separation \\
\end{tabular}
\end{ruledtabular}
  \caption{\label{tab:resolutions}Tracking momentum and impact parameter resolution, tracking efficiency, calorimetry resolution, and PID performance that are used as input for \textsc{Delphes} fast simulations. These parameters are partially based on Ref.~\cite{EICHandbook}.}
\end{table}

\subsection{Jet kinematics}
\label{sec:jetkinematics}

Jets are reconstructed with the anti-$k_{T}$ algorithm~\cite{Cacciari:2008gp} and $R\!=\!1.0$ as implemented in \textsc{Fastjet}~\cite{Cacciari:2011ma}. The choice of $R\!=\!1.0$ follows the HERA experiments, which showed this definition minimized hadronization corrections ~\cite{Newman:2013ada}. Jets are defined both ``at the generator level'' and at the ``reconstructed level''; the input for the generator  level are final-state particles in \textsc{Pythia8} (excluding neutrinos), whereas the input for the reconstructed level are particle-flow objects from \textsc{Delphes}. Reconstructed jets are matched to generated jets with an angular distance selection of $\Delta R = \sqrt{(\phi_{jet}^{gen}- \phi_{jet}^{reco})^{2} + (\eta_{jet}^{gen}- \eta_{jet}^{reco})^{2}}  <0.5$ (half the radius parameter). The requirement that an electron-proton collision produce a reconstructed jet within the tracking fiducial region, $|\eta|<3.0$, is 95\% efficient on CC DIS events.

\begin{figure}
    \centering
        \vspace*{-1.2cm}
    \includegraphics[width=1.0\columnwidth]{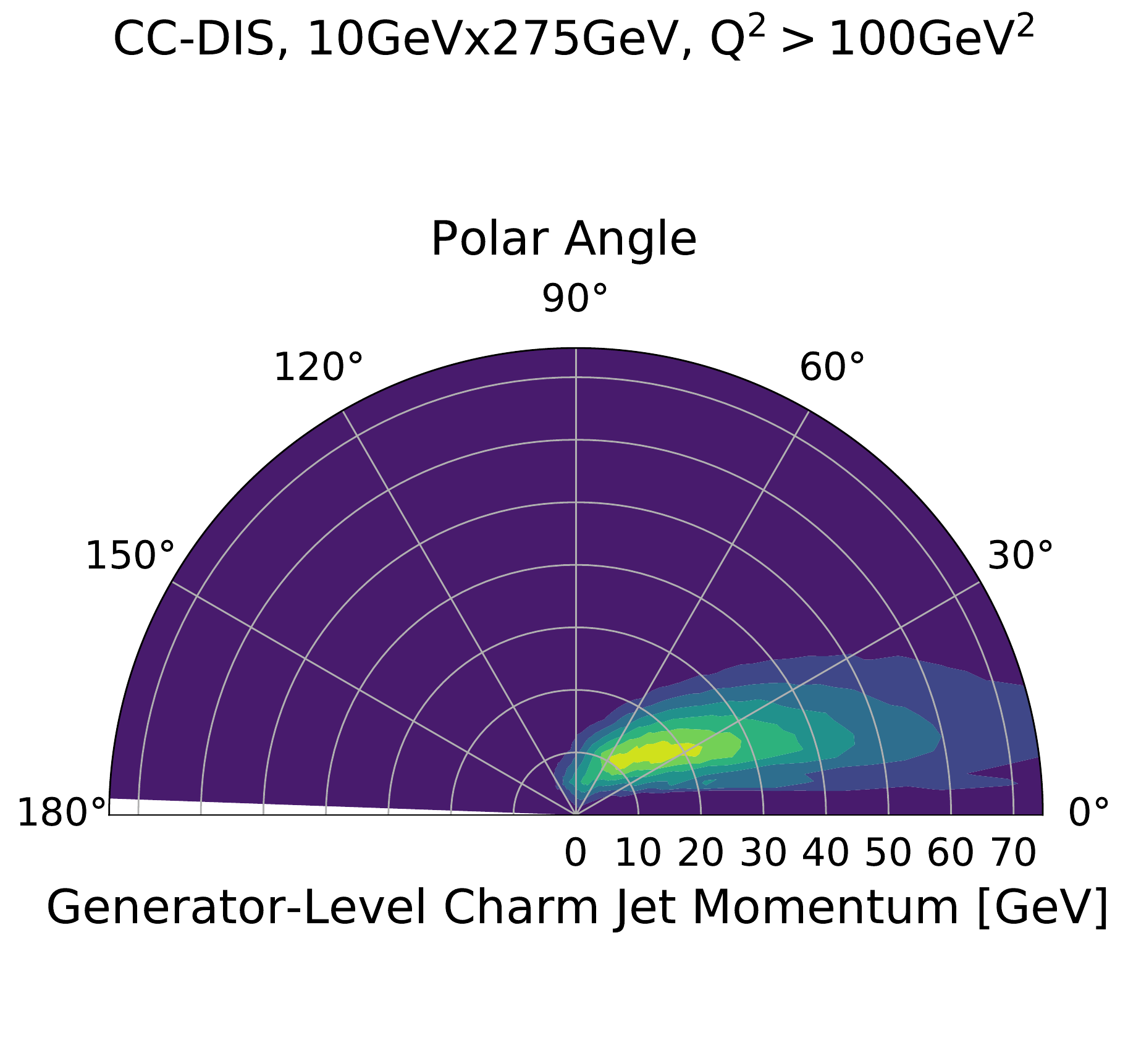}
            \vspace*{-1.2cm}
\caption{\label{fig:charm_jet_kinematics}The kinematics (momentum, $p$, and polar angle, $\theta$, with respect to the direction of the hadron beam) of generated charm jets in CC DIS with $Q^{2}>100$ GeV$^{2}$. The jets are clustered with the anti-$k_{\mathrm{T}}$ algorithm with $R=1.0$.  }
\end{figure}

 Figure~\ref{fig:charm_jet_kinematics} shows the kinematics\footnote{We follow the HERA convention to define the coordinate system: the $z$-direction is defined along the beam axis and the electron beam goes towards negative $z$. The polar angle $\theta$ is defined with respect to the proton direction.} of charm jets, which lie prominently at low angles to the positive $z$-axis ($\eta \approx 1.3$) with momenta of $p \approx 15\mathrm{~GeV}$. However, a significant fraction of jets are produced at even shallower angles up to $\eta \approx 3$; accounting for the large radius parameter of these jets, this implies that efficient reconstruction and tagging of charm jets will require tracking and calorimeter coverage out to $\eta=3.5-4.0$, consistent with the baseline EIC detector described above.

\begin{figure}
    \centering
    \includegraphics[width=1.0\columnwidth]{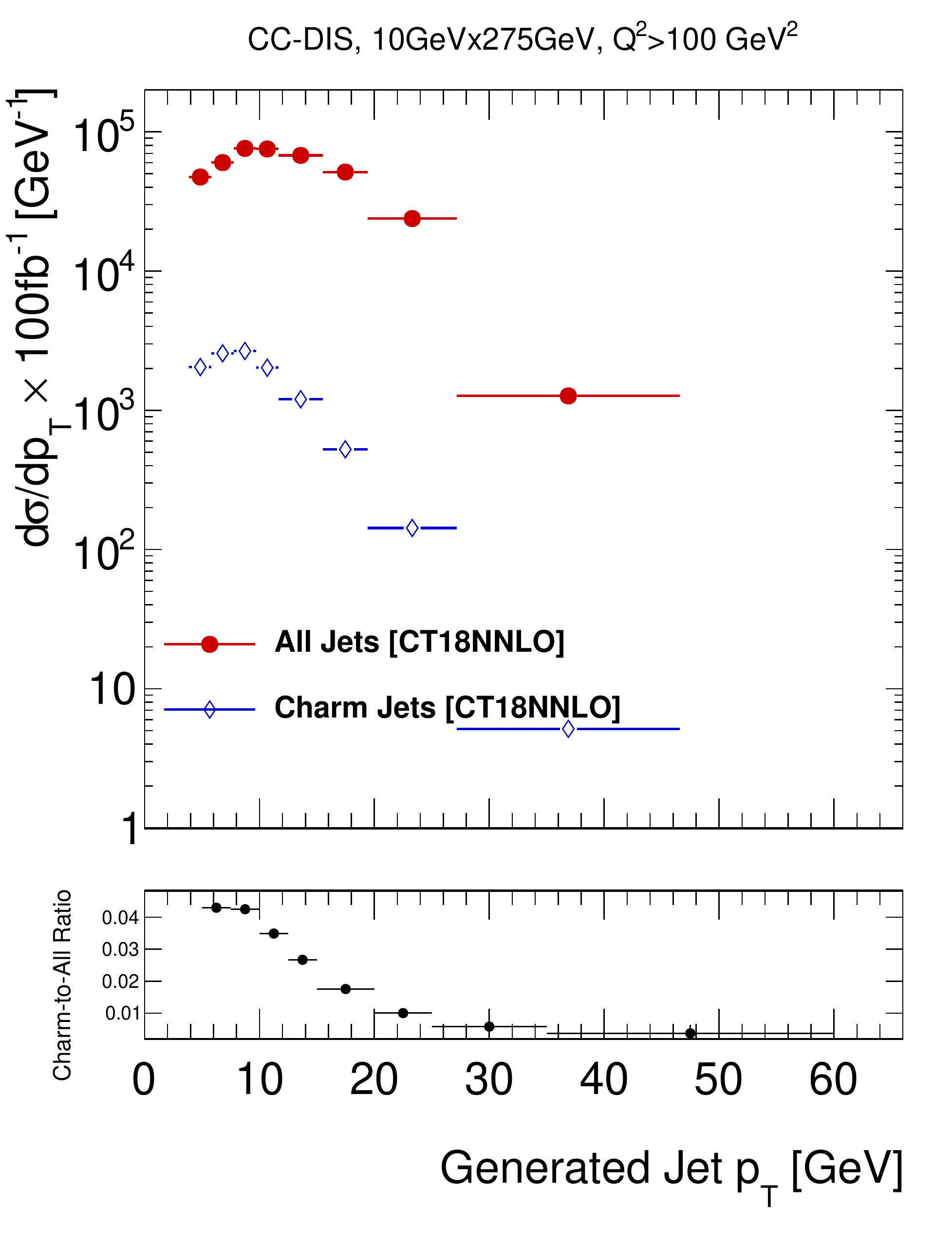}
\caption{\label{fig:jet_production}Inclusive and charm-jet production in charged-current DIS at generator level. The jets are reconstructed with the anti-$k_{\mathrm{T}}$ algorithm with $R=1.0$.}
\end{figure}

The inclusive and charm-jet \pT~cross sections are shown in Figure~\ref{fig:jet_production}. The ratio of charm-to-inclusive cross section is about 3.5\% at $\pT\! =\! 10$ GeV and it decreases to less than 0.5\% at 40 GeV. The jet \pT~is correlated with $x$, so this decrease reflects the faster drop of the strange PDF with respect to the valence quark PDFs. 

\subsection{Event selection, jet and missing-energy performance}
\label{sec:event_selection}

Following HERA measurements~\cite{Abramowicz:2015mha}, the tagging of charged-current DIS events is obtained by requiring large missing-transverse energy (\met), which is defined as the magnitude of the vector sum of the transverse momenta of all \textsc{Delphes} particle-flow objects. It is defined at the generator level in a similar way but using all stable generated particles, or equivalently, by neutrinos. 

\begin{figure}
\centering
\includegraphics[width=.99\columnwidth]{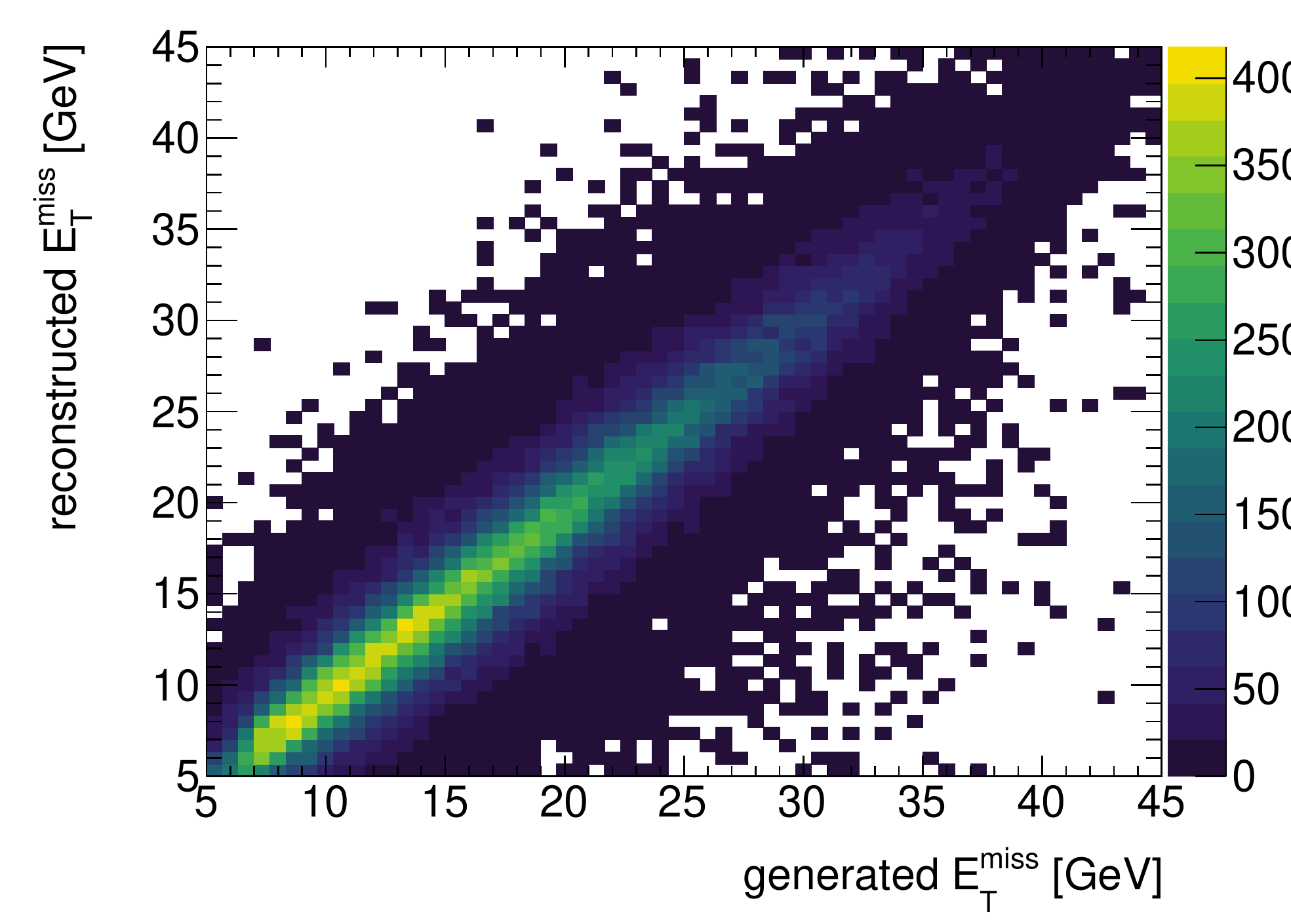}
\caption{\label{fig:metperformance} Missing-transverse energy (\met) response matrix for charged-current DIS events. The missing energy is reconstructed using \textsc{Delphes} particle-flow objects.}
\end{figure}

Figure~\ref{fig:metperformance} shows the \met\ performance obtained with the baseline parameters. The relative \met\ resolution ranges from 20\% (23\%) at 10 GeV to 6\% (11\%) at 40 GeV, defined with a Gaussian fit (standard deviation). The difference between the relative resolutions obtained with a Gaussian fit and the standard deviation reflect the tails of the response, which primarily come from losses due to thresholds in tracking and calorimetry. 

The relative jet \pT~resolution ranges from 18\% (21\%) at 10 GeV to 7\% (12\%) at 40 GeV, defined with a Gaussian fit (standard deviation). We have studied what happens to jets and \met\ in the case that the hadronic calorimeter provides less or no coverage in the barrel region ($|\eta|<1.0$), such as in the BEAST~\cite{Aschenauer:2014cki} or JLEIC~\cite{Morozov:2019uza} detector designs. This leads to a significant and asymmetric tail in the jet and \met\ response. That feature would complicate future unfolding procedures as well as background rejection for photoproduction and NC DIS. Our results agree with Page et al.~\cite{Page:2019gbf}, who reported that the lack of barrel hadronic calorimeter leads to a severe degradation of jet performance. 

We use the Jacquet-Blondel method~\cite{Amaldi:1979yh} to reconstruct the event kinematics: the event inelasticity is reconstructed as: $y_{JB} = \sum_{i}(E_{i}-p_{z,i})/E_e$, where the sum runs over all particles in the event (particle-flow objects) and $E_{e}$ is the electron beam energy; the transfer-momentum squared is $Q_{JB}^{2} = (\mathrm{\met})^{2}/(1-y_{JB}) $ and Bjorken $x$ is $x_{JB}= Q^{2}_{JB}/sy_{JB}$, where $s=4E_{e}E_{p}$ and $E_{e}$ ($E_{p}$) is the energy of the electron (proton) beam. We compute the ``bin-survival probability'' defined as $p^{i} = \left(N_{gen}^{i}-N_{out}^{i})/(N_{gen}^{i}-N_{out}^{i}+N_{in}^{i}\right)$, where $N_{gen}$ is the number of events generated in a bin $i$; $N_{out}$ is the number of events generated in bin $i$ but reconstructed in bin $j\neq i$; and $N_{in}$ is the number of events generated in a bin $j\neq i$ but reconstructed in bin $i$.

\begin{figure}
\centering
\includegraphics[width=1.0\columnwidth]{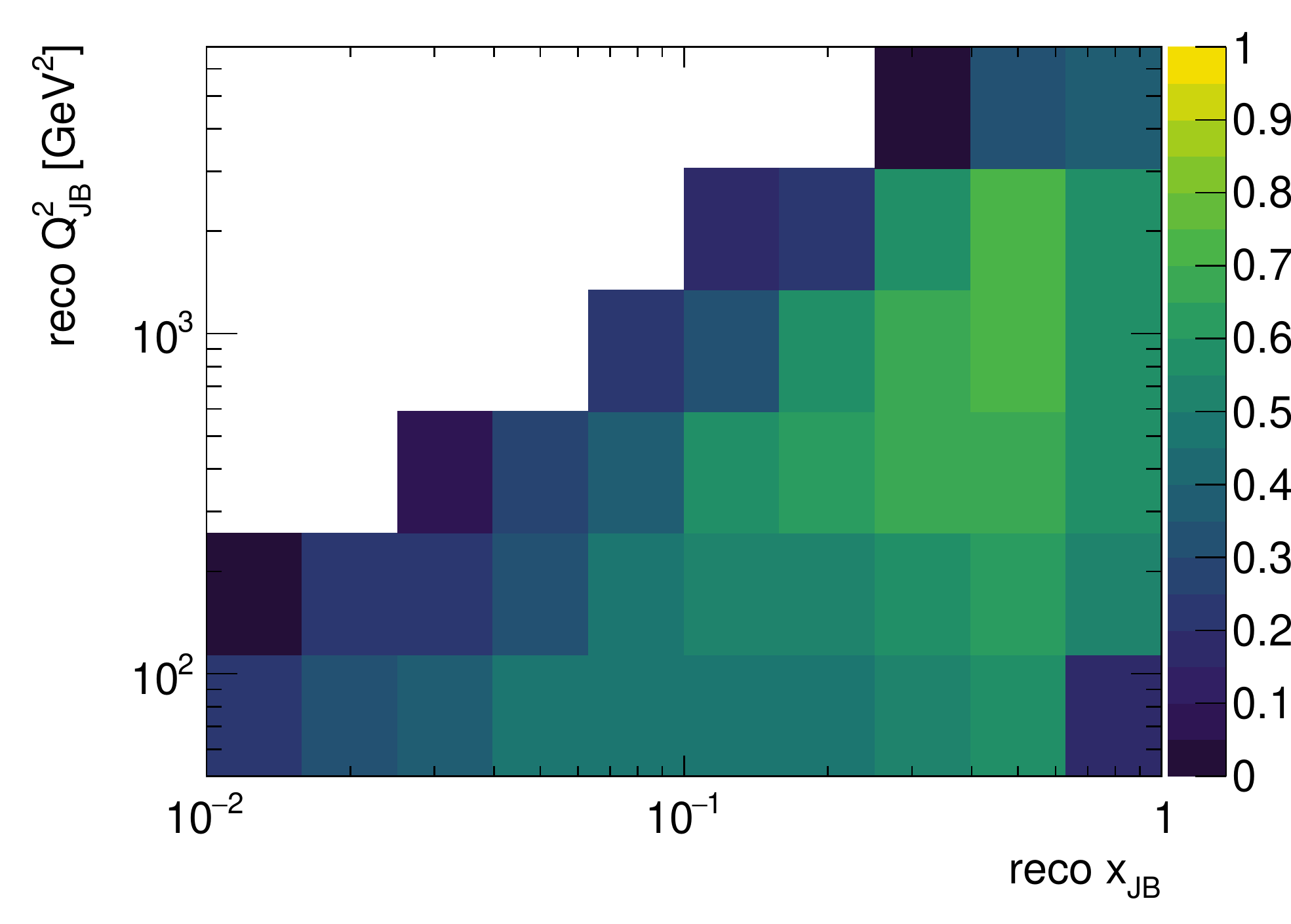}
\caption{\label{fig:purity} Bin survival probability obtained using the Jacquet-Blondel method in charged-current DIS events.}
\end{figure}

Figure~\ref{fig:purity} shows the two-dimensional bin-survival probabilities, which are about 70$\%$ or better for a large region at high $Q^{2}$ and $x$. Similar results were presented by Aschenauer et al.~\cite{Aschenauer:2013iia} using the BEAST detector design parameters. This level of bin-survival probability would enable a controlled unfolding procedure in two dimensions ($x$ and $Q^{2}$ or $x$ and $\pTjet$). In this work, we focus on one-dimensional distributions (either $x$, or the $\pTjet$ spectrum) and leave detailed unfolding studies to future work. 

We select events with \met\ $> 10$ GeV. From our simulation of CC DIS events with a reconstructed fiducial jet, this requirement is 87\% efficient; we find it to be 75\% efficient on events that contain a reconstructed, truth-matched charm jet. Background from photo-production and NC DIS is suppressed by the \met\ selection and can be further suppressed by topological cuts, far-forward tagging of electrons, and kinematic constrains, as was done by the HERA experiments~\cite{Abramowicz:2015mha}. In the following, we assume that these backgrounds can be made negligible with little impact on CC DIS selection efficiency. 

\section{Charm Jet Tagging}
\label{sec:tagging}

\subsection{Displaced Track Counting}
\label{sec:ip3d}

After jet reconstruction and \met\ selection, we use a charm-jet tagging algorithm that employs the counting of high-impact-parameter tracks. The $c\tau$ of charm hadrons varies between about 0.2-0.5$\mathrm{~mm}$~\cite{Tanabashi:2018oca}, and, for typical charm jets produced at EIC energies, this results in flight lengths of up to a few millimeters from the interaction point. The decay of the charm hadron can result in one or more tracks whose impact parameter is significantly displaced from the interaction point.

\def\IPTD{\ensuremath{\mathrm{IP_{3D}}}}
\def\sIPTD{\ensuremath{\mathrm{sIP_{3D}}}}

\begin{figure}
    \centering
    \includegraphics[width=1.0\columnwidth]{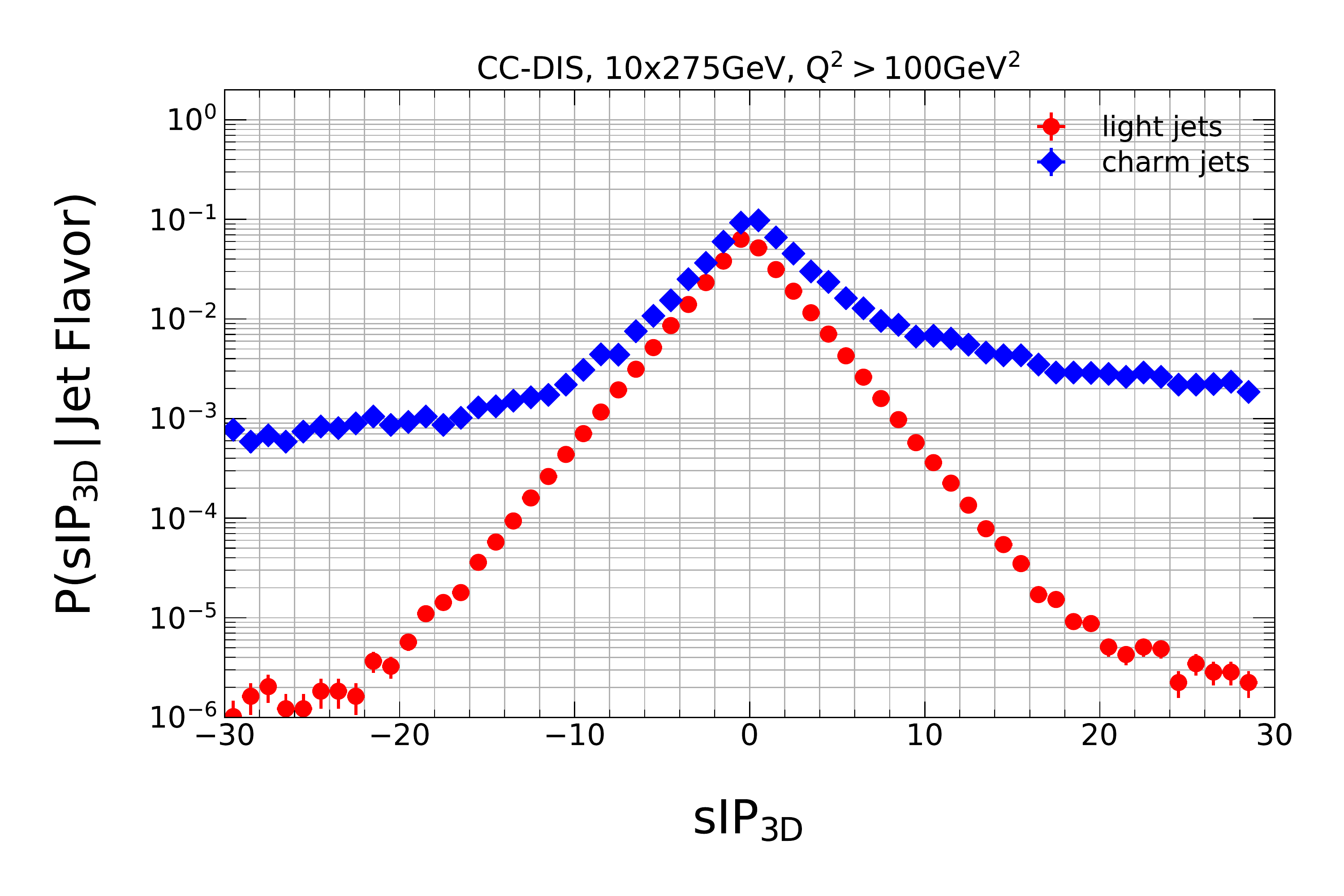}
\caption{\label{fig:track_ip}Signed impact parameter significance, $\sIPTD$, probability distribution for light and charm jets.}
\end{figure}

We match tracks to a jet and compute the distance of closest approach to the interaction point in the $x-y$ plane ($d_0$) and along the $z$-axis ($z_0$). The 3-D impact parameter significance is then defined as $\IPTD=\sqrt{(d_0/\sigma_{d_0})^2+(z_0/\sigma_{z_0})^2}$. We assume a resolution of $\sigma_{z_0} = \sigma_{d_0} = 20\mathrm{\mu m}$.  The signed impact parameter, $\sIPTD$ (Fig.~\ref{fig:track_ip}), is obtained by multiplying $\IPTD$ with the sign of the product $\vec{p}_j \cdot \vec{r}_{\mathrm{track}}$, where $\vec{p}_j$ is the parent jet momentum and $\vec{r}_{\mathrm{track}}$ is a vector that points from the interaction point to the point of closest approach on the track. 

\begin{figure}
    \centering
    \includegraphics[width=1.0\columnwidth]{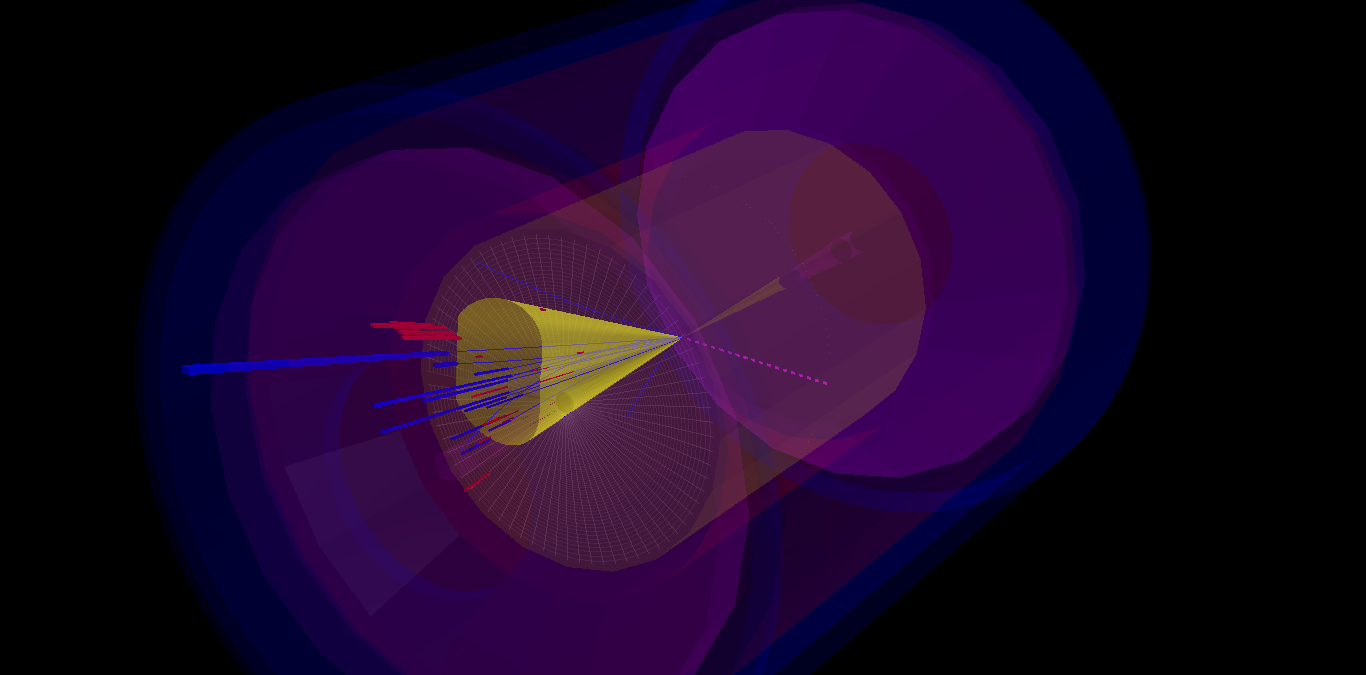}
    \\
    \includegraphics[width=1.0\columnwidth]{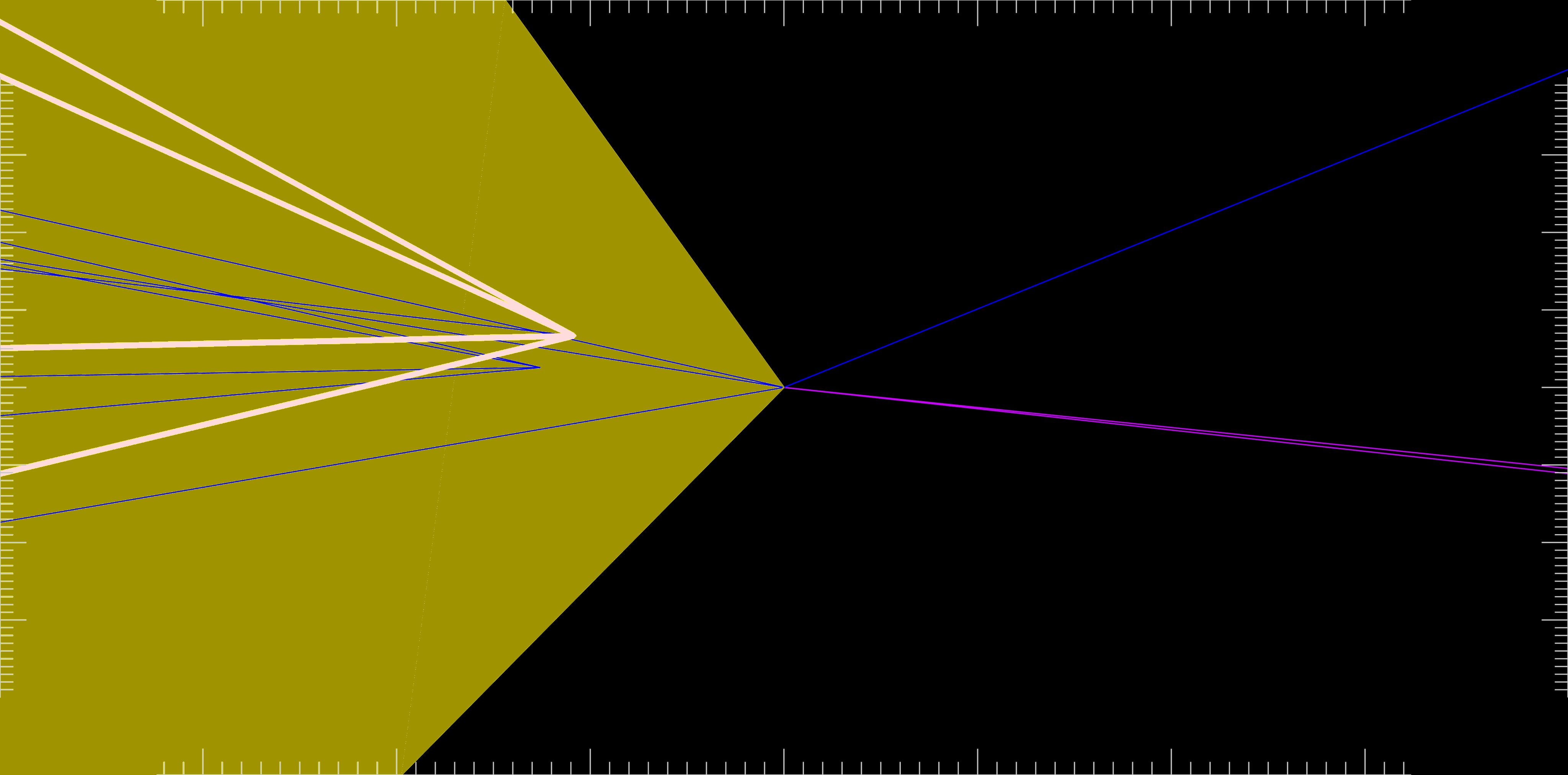}
    
\caption{\label{fig:display}A pair of event displays of a single CC DIS event simulated with \textsc{Pythia8} and reconstructed with \textsc{Delphes}. A reconstructed jet is represented as a yellow cone; blue bars are hadronic calorimeter energy deposits, and red bars are electromagnetic calorimeter energy deposits. Tracks are indicated by blue lines; the yellow-highlighted tracks originate from a displaced decay vertex. The zoomed-in view (bottom) shows these tracks and vertices.}
\end{figure}

A basic optimization of the tagger parameters was performed by maximizing the significance of the background-subtracted charm jet yield assuming a target integrated luminosity of $100\mathrm{~fb^{-1}}$. This leads to the requirements of $\ge 2$ tracks, each of which satisfies $\pT^{track}>0.5\mathrm{~GeV}$; $\sIPTD>3.00$; and $\sqrt{d_0^2 + z_0^2}<3\mathrm{mm}$. A jet meeting these criteria is referred to as ``tagged.'' This approach selects both long-lived charm and bottom jets. An example of such a charm-tagged jet is shown in Fig.~\ref{fig:display}.

\begin{figure}
    \centering
    \includegraphics[width=\columnwidth]{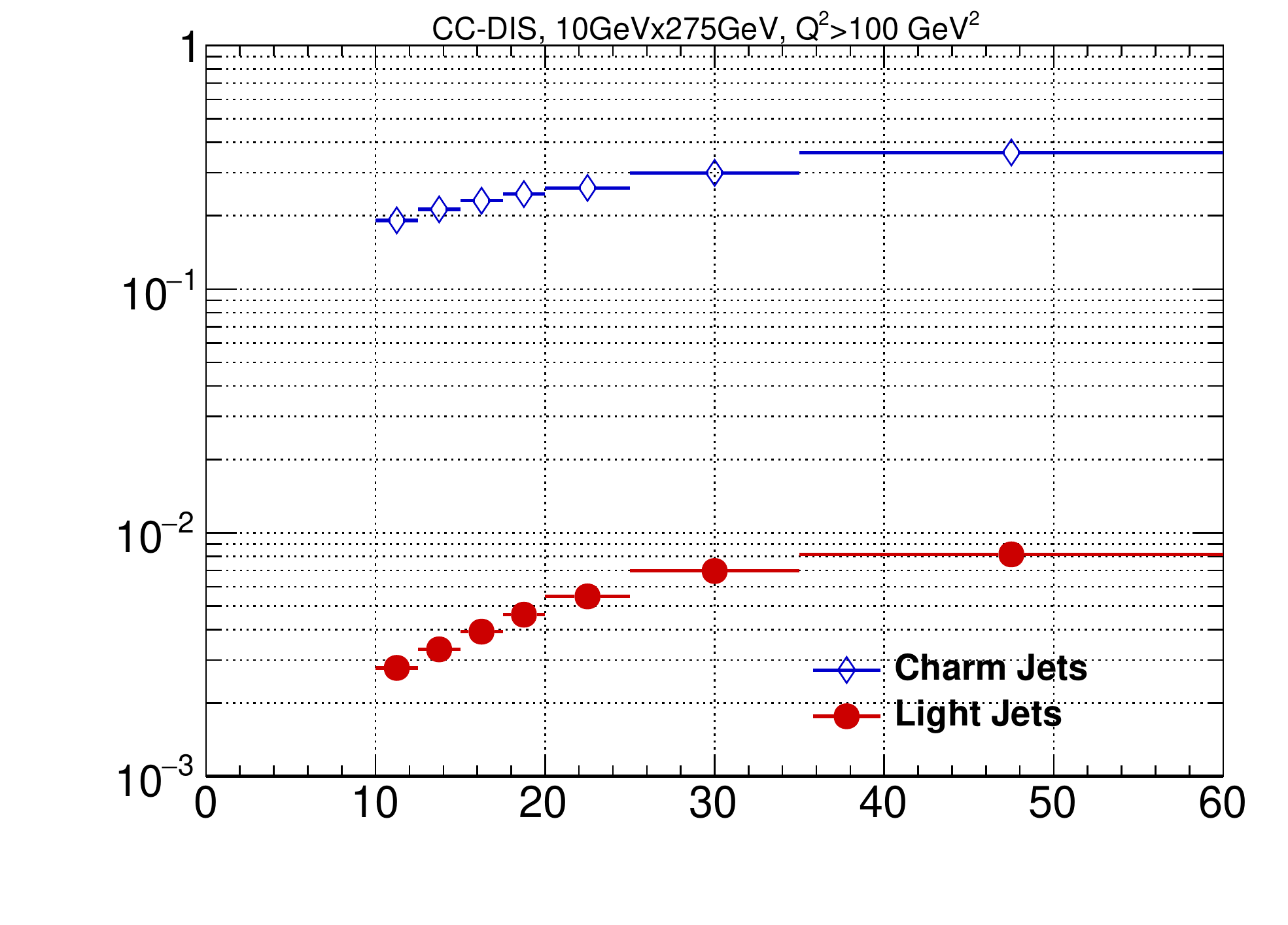}
\caption{\label{fig:charm_tag_eff}Jet tagging efficiency for light- and charm jets as a function of the jet \pT. The high-impact parameter track-counting approach is used to obtain these results.}
\end{figure}

The tagging efficiency is defined by identifying all jets matched in the simulation to either a bottom hadron, charm hadron, or light hadron (in that hierarchy), and then determining the number relative to each population that additionally pass the tagging requirement. The performance is summarized in Fig.~\ref{fig:charm_tag_eff}. For charm jets with $\pTjet>10\mathrm{~GeV}$, this basic approach leads to charm-jet efficiencies ranging between 20-40\%  and light-jet efficiencies between $(0.2-1)\%$. The average charm (light) jet efficiency in this jet $\pT$ range is 20\% (0.4\%). These efficiencies lead to roughly equal populations of charm and light jets in the tagged jet sample. When reporting the uncertainty on charm jet yields ($\sigma_{s}$), we assume that the light-jet component ($n_b$) can be subtracted from the tagged jet population ($N$), such that the background-subtracted uncertainty is given by $\sigma_{s} = \sqrt{N+n_b}$.

As an example of the effect of detector performance on charm-jet tagging efficiency, we degrade the impact parameter resolution along the $z$ axis from $20\mathrm{~\mu m}$ to $100\mathrm{~\mu m}$. We then re-optimize the tagging approach to see if compensation for the degradation is possible by re-tuning the requirements on the hyper-parameters. The average charm (light) jet tagging efficiency, with re-optimization, moves from 20\% (0.4\%) to 14\% (0.4\%). The re-optimization generally maintains all hyper-parameter requirements except that on the minimum flight significance, which is loosened to $2.75\sigma$ to maintain background rejection and signal efficiency as the flight significance distribution is diluted by the increased resolution. This change in performance represents a 30\% loss in charm-jet tagged yield as a result of this degradation of $z_0$ resolution, while the light-jet background is unchanged. If we additionally degrade the $d_0$ resolution to $100\mathrm{\mu m}$, we observe a further dilution-induced loosening of the flight significance requirement and a corresponding decline in re-optimized tagging efficiency to 8.0\% (for a light-jet efficiency of 0.6\%). This would represent an overall loss of 60\% of tagged charm jets from the baseline scenario, as well as a further increase in light-jet contamination. 

We also assessed a more optimistic scenario in which the tracking system permits an improved impact parameter resolution, $\sigma_{d_0}=\sigma_{z_0}=10\mathrm{\mu m}$, over the EIC baseline. A re-optimization under this case results in maintaining the same hyper-parameters as in the baseline scenario. However, due to the improved resolution the charm jet efficiency increases to 26\% while the light-jet efficiency remains at 0.4\%. For the same light-jet background, this represents a 30\% gain in charm jets.

\subsection{Single-Track PID Jet-Tagging Approaches}
\label{sec:pidtagging}

Dedicated PID approaches are anticipated as part of the baseline EIC detector. For example, calorimeter-only methodologies (ECAL/HCAL) can be used to separate electrons from pions or other hadrons, as well as using responses from other sub-components like Cherenkov, preshower, or transition-radiation detectors.  An additional dedicated muon system could be employed to separate muons especially from pions; a Cherenkov radiation detector could be used to separate kaons from pions. We considered the potential of such systems for charm-jet tagging.

We studied the impact on charm jet tagging if we employed searches in jets for single, high-impact parameter, well-identified kaons, muons, or electrons. Since about 80\% of actual charm jets reconstructed in the detector simulation are untagged by the approach in Section~\ref{sec:ip3d}, we explored the additional tagging efficiency that might be recovered.

We consider only tracks with $\pT>1\mathrm{~GeV}$ and with $\mathrm{sIP_{3D}} \ge 3$. We emulate a future PID approach or system by selecting true charged kaons, electrons, pions, or muons contained in a reconstructed jet and applying the following conservative efficiencies/mis-identification rates:

\begin{itemize}

    \item 90\% kaon identification efficiency and a 0.44\% pion mis-identification rate ($3\sigma$ kaon-pion separation). This is consistent with the EIC detector baseline.
    \item 90\% electron identification efficiency and a 2\% pion mis-identification rate; this corresponds to a rejection factor of 50 for pions, or a $2.4\sigma$ electron-pion separation.
    \item 95\% muon identification efficiency and a 5.4\% pion mis-identification rate ($2\sigma$ muon-pion separation).
\end{itemize}

Using these approaches on the charm jets un-tagged by the $\sIPTD$ approach, we found an additional 2\% charm-jet tagging efficiency gain using just electrons; an additional 3\% efficiency gain using just muons; and an additional 6\% efficiency gain using just kaons. Combining all three in a logical "OR" resulted in tagging an additional 11\% of charm jets previously un-tagged by the track-counting approach. Using all methods together brought the total charm-jet tagging efficiency to 31\%. We note that we did not optimize the hyper-parameters (track momentum, $\sIPTD$) for this PID-based study, but applied reasonable values given the event kinematics. 

We observe that these single-track PID approaches have comparable light-jet mis-identification rates to the high-impact-parameter track approach. The mis-tagging rate of light jets using the described methods to select single electrons, muons, or kaons was $0.06\%$, $0.1\%$, and $0.1\%$, respectively. In combination with the high-impact-parameter track-counting approach, we would expect to see an overall increase in the significance of the charm jet yield. Our PID performance assumptions are conservative, offering an expectation of strong performance and sensitivity gains by combining PID-based and other approaches if improvements over our assumptions are achieved.

\section{Sensitivity to Strangeness}
\label{sec:strange}

\def\rslow{\hbox{\textsc{Rs{-}Low}}}
\def\rshigh{\hbox{\textsc{Rs{-}High}}}

At leading order in $\alpha_s$, final-state charm production in CC DIS is
driven by the flavor excitation process shown in Fig.~\ref{fig:feynman_diagram},
in which an initial-state (anti)strange quark absorbs
a $W^{\mp}$ boson to excite (anti)charm~\cite{Aivazis:1990pe}. For this reason, leading-order
charm-jet production has direct sensitivity to the proton's strange-quark
content. In the present analysis, we examine the event-level impact of
varying the input strangeness within a set of extremal bounds determined
within the CT18 global fit \cite{Hou:2019efy}.

\subsection{Theory inputs: extreme \texorpdfstring{$R_s$}{Rs} scenarios}
\label{sec:Rs}

\begin{figure}
    \centering
    \includegraphics[width=1.0\columnwidth]{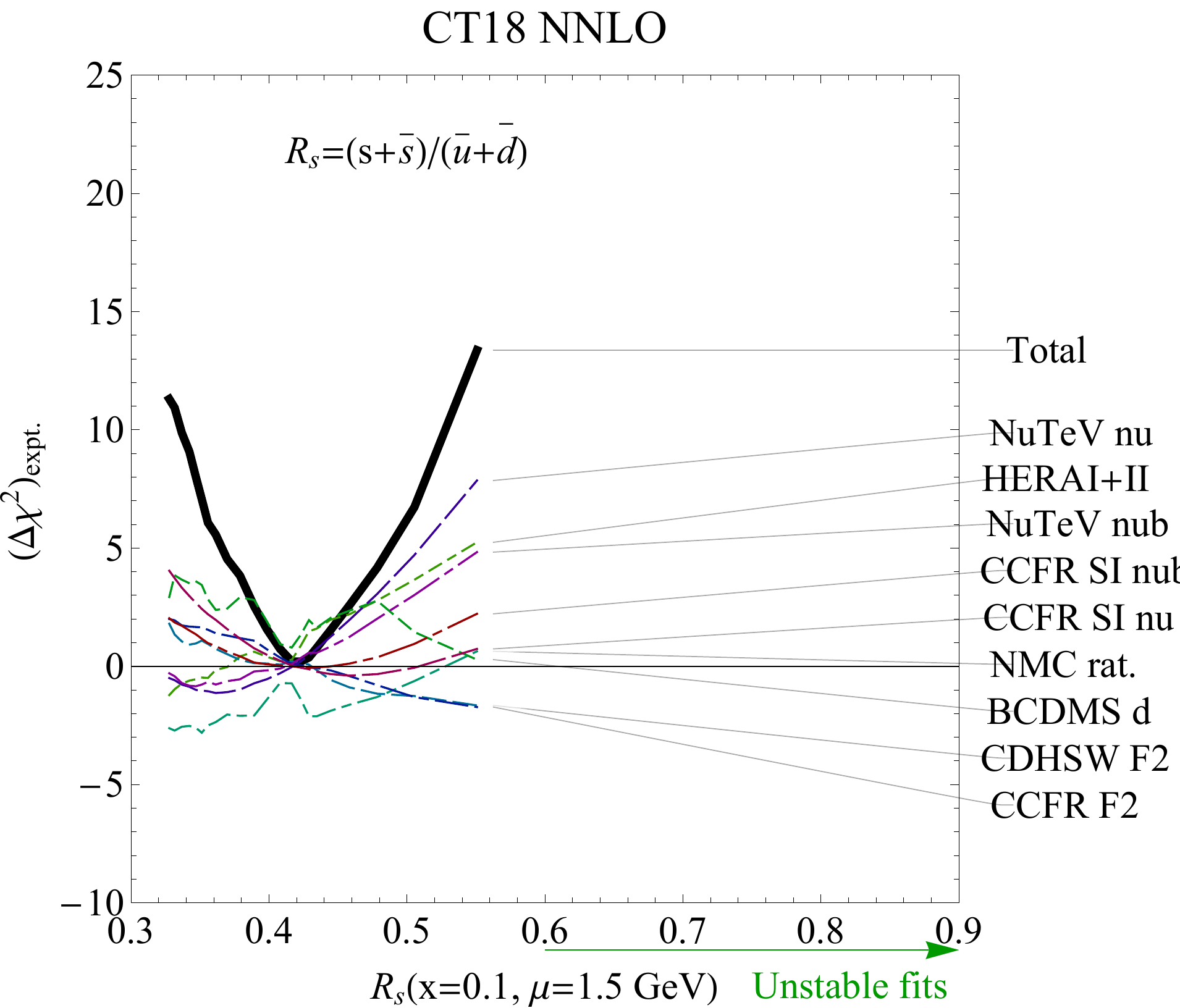} \\
    \includegraphics[width=1.0\columnwidth]{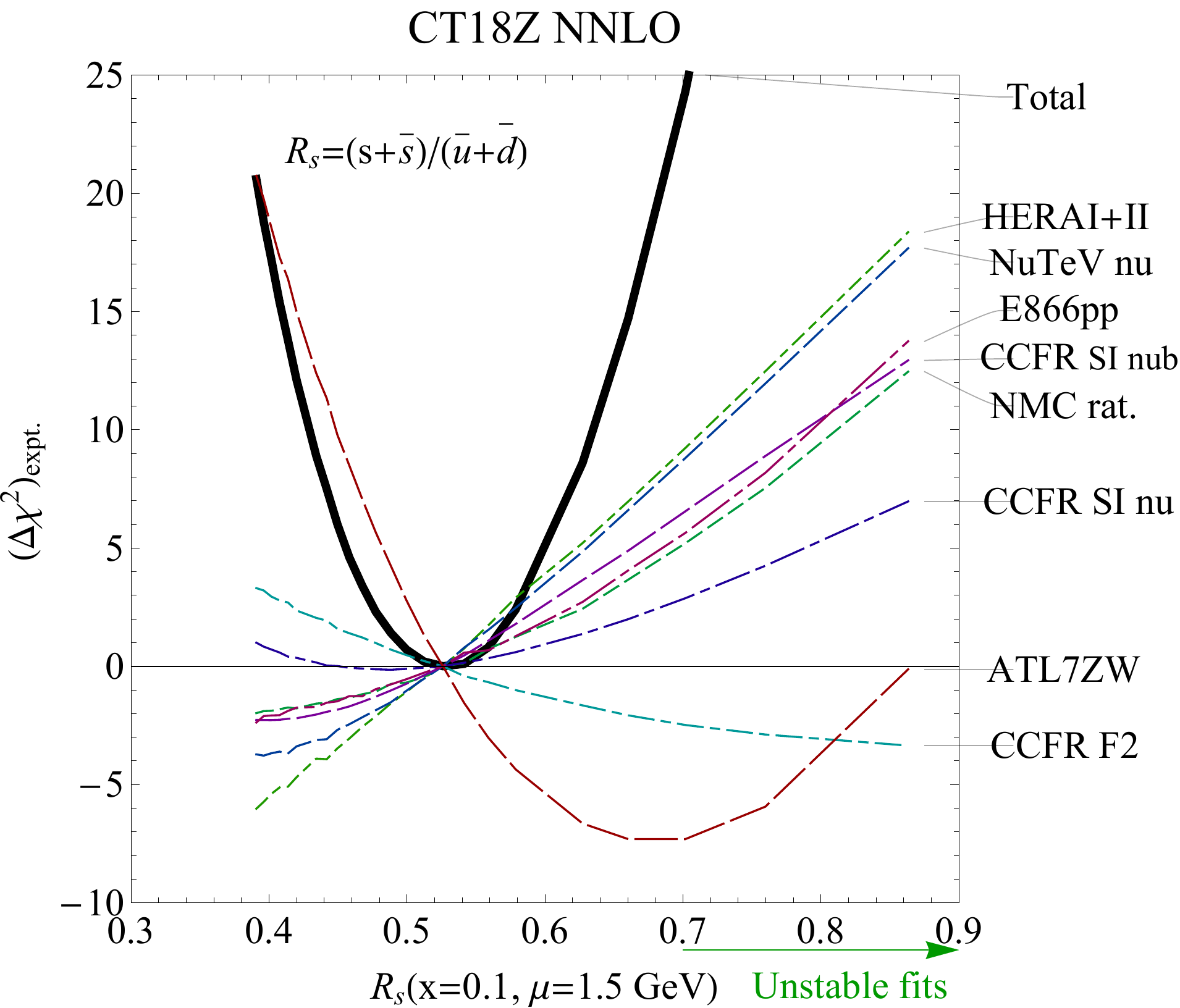}
\caption{To gauge the event-level sensitivity of charm-jet production, we perform simulations
        with two extreme inputs for the behavior of the strangeness suppression ratio,
        $R_s$, as defined in Eq.~(\ref{eq:Rs}), taken from the recent
        CT18 NNLO global PDF analysis \cite{Hou:2019efy}. The upper panel
        corresponds to a Lagrange Multiplier (LM) scan over values of $R_s$ at high $x\!=\!0.1$ in the primary CT18
        baseline fit, while the lower panel was obtained for the alternative
        CT18Z fit, which included the ATLAS 7 TeV $W/Z$ data (ATL7ZW) \cite{Aaboud:2016btc},
        in addition to a number of other modifications. In both panels, the PDF
        scale is the factorization scale, $Q = \mu = 1.5\mathrm{~GeV}$.
        }
\label{fig:Rs}
\end{figure}

As a proxy for different behaviors of the light-quark sea, we extremize
inputs for the high-$x$ behavior of the strange suppression factor,
defined in Eq.~(\ref{eq:Rs}). 
This quantity has in general received significant attention, as
it quantifies the extent to which hadronic-scale QCD interactions
lead to the violation of the flavor symmetry commonly assumed
in the earliest PDF analyses, {\it i.e.}, $s = \bar{s} = \bar{u} = \bar{d}$.
Historically, PDF fits assumed a suppressed intrinsic strangeness
by fixing $R_s = 0.5$, such that the $x$ dependence of the proton's
$s$-PDF was entirely determined by that of the $\bar{u},\,\bar{d}$
anti-quark PDFs. As noted in Sec.~\ref{sec:intro}, such choices
have primarily been made given the sparsity of data with direct
sensitivity to nucleon strangeness, including semi-inclusive
kaon production and dimuon production in neutrino-nucleus DIS.
In fact, even modern PDF fits that exclude this latter data,
including CJ15, do not actively fit nucleon strangeness,
instead taking $R_s = \kappa = 0.4$ \cite{Accardi:2016qay}.

At the same time, an independent strange component of the nucleon wave
function has long been the subject of modeling efforts in
non-perturbative QCD \cite{Signal:1987gz,Chang:2011vx,Hobbs:2014lea,Hobbs:2016xlg},
lattice studies \cite{Junnarkar:2013ac,Liang:2019xdx}, and dedicated global PDF
analyses \cite{Lai:2007dq,Sato:2019yez}.
For these reasons, as well as the importance of detailed knowledge of the
nucleon sea's flavor structure for precision phenomenology at hadron
colliders like the LHC, PDF fitting efforts like the CTEQ-TEA (CT)
Collaboration have had a sustained interest in the constraints high-energy
data place on $s(x,Q)$ and $R_s(x,Q)$. The most recent iteration of
PDF fits developed within the CT global analysis framework --- CT18 ---
were recently released in Ref.~\cite{Hou:2019efy}. This latest fit examined
implications of the recent LHC Run-1 data for the $s$-PDF, which extended the sensitivity of the global data set beyond that
driven by the legacy data included in older fits; among these legacy data
are the fixed-target neutrino DIS experiments such as CCFR \cite{Goncharov:2001qe}
and NuTeV \cite{Mason:2006qa}, which still provide the dominant PDF pulls
in the strange sector.

A particular subtlety explored in Ref.~\cite{Hou:2019efy} is the
theoretical description of the recent ATLAS 7 TeV $W/Z$ production
data \cite{Aaboud:2016btc}, which generally prefer an enlarged
strange PDF and comparatively larger value of $R_s$.
For instance, at more intermediate $x\!=\! 0.023$ and $Q^2=1.9\,\mathrm{GeV}^2$,
the ATLAS collaboration reported a value of
$R_s = 1.13 \pm 0.05 (\mathrm{exp}) \pm 0.02 (\mathrm{mod})^{+0.01}_{-0.06} (\mathrm{par})$
based upon an internal fit, suggesting unsuppressed strangeness along the lines
of the earlier 2012 ATLAS result \cite{Aad:2012sb}. The preference of the ATLAS $W/Z$ data
for enlarged strangeness was confirmed by the CT18 global analysis as well as
other recent studies \cite{Ball:2017nwa,Thorne:2019mpt,Kusina:2020lyz}.
A detailed discussion of these data and challenges associated with their
theoretical description is presented in App.~A of
Ref.~\cite{Hou:2019efy}. 
Ultimately, the ATLAS 7 TeV data were
not treated in the CT18 main fit, but rather in alternative fits, CT18A/Z,
which either included these data on top of the other sets fitted in CT18 (CT18A) or
included them along with several other alternate choices for theory settings
and data set selections (CT18Z).
We point out that the ATLAS 7 TeV $W/Z$ data have attracted significant interest
from a number of other PDF fitting groups, including a recent study~\cite{Faura:2020oom}, who
considered them in the context of several neutrino-scattering experiments, such
as the NuTeV dimuon-production measurements, as well as NOMAD; in this context,
Ref.~\cite{Faura:2020oom} reports $\chi^2/N_\mathit{pt}\! =\! 1.61$ for the
$N_\mathit{pt}\! =\! 61$ ATLAS $W/Z$ data set, suggesting a decent description
of these data can be obtained in a fit that also includes the neutrino information
with an intermediate strangeness of $R_s = 0.71 \pm 0.10$ for $x=0.023$ and $Q=1.6$ GeV.

This brings us to the central question of this analysis: {\it can high-precision
charm-tagged data obtained from CC DIS jet production at the EIC help resolve the
size and $x$ dependence of the strange-quark sea}? If possible,
the EIC would then play a pivotal role via the charm-jet CC DIS channel in 
further navigating
any potential tensions between the pulls of the $\nu A$ DIS and $W/Z$
hadroproduction data on $R_s$ and nucleon strangeness, 
and providing
critical added constraints to reduce remaining PDF uncertainties. For the present
feasibility study, we explore this question by examining the event-level
discriminating power of CC DIS jet simulations upon widely-separated theory inputs
for $R_s$. Given the especially strong resolving power of CC DIS jet measurements
at high $x$, we therefore examine whether such hypothetical data might
be sensitive to two extreme sets for different behavior of $R_s$ at
$x=0.1$ near the non-perturbative starting scale. We note that the current
study concentrates on total strangeness; in CT18, which used
$s\!=\!\bar{s}$, this is therefore $2s\! =\! s\!+\!\bar{s}$. In principle,
an EIC with positron beams could provide an advantageous setting to
test $s\! \neq \!\bar{s}$. We leave this question to future work.
\newcommand\T{\rule{0pt}{14pt}}       %
\newcommand\B{\rule[-8pt]{0pt}{0pt}} %
\newcommand\tb{\rule{0pt}{11pt}\rule[-6pt]{0pt}{0pt}} %
\newcommand\error[2]{\genfrac{}{}{0pt}{}{#2}{#1}}
\newcommand\brule[1]{\textcolor{#1}{\rule{1.5cm}{0.2cm}}}
\definecolor{darkgreen}{rgb}{0,0.67,0}
\definecolor{purple}{rgb}{0.5,0.0,0.5}
\begin{table}[tbh]
    \centering
\begin{tabular}{|l|c|c|}
\hline 
\T \B PDF Set & $\kappa(Q)=\int R_{s}(x,Q)$ & color\tabularnewline
\hline 
\hline 
\tb \textcolor{red}{\bf \rslow{}} & $0.37$ & \brule{red}\tabularnewline
\hline 
\tb CJ15nlo & $0.43\:$$\pm0.01$ & \brule{black}\tabularnewline
\hline 
\tb CT18NNLO & $0.44\:$$\error{-0.11}{+0.15}$ & \brule{brown}\tabularnewline
\hline 
\tb MSTW2008nnlo68cl & $0.48\:$$\error{-0.03}{+0.02}$ & \brule{blue}\tabularnewline
\hline 
\tb EPPS16\_CT14nlo\_Pb208 & $0.49\:$$\error{-0.19}{+0.21}$ & \brule{orange}\tabularnewline
\hline 
\tb nCTEQ15FullNuc\_208\_82 & $0.50\:$$\pm0.01$ & \brule{green}\tabularnewline
\hline 
\tb HERAPDF20\_NLO\_VAR & $0.57\:$$\error{-0.42}{+0.19}$ & \brule{darkgreen}\tabularnewline
\hline 
\tb NNPDF31\_nnlo\_as\_0118 & $0.59\:$$\error{-0.30}{+0.30}$ & \brule{magenta}\tabularnewline
\hline 
\tb CT18A NNLO & $0.63\:$$\error{-0.16}{+0.23}$ & \brule{purple}\tabularnewline
\hline 
\tb  \textcolor{red}{\bf \rshigh{}} & $0.96$ & \brule{red}\tabularnewline
\hline 
\end{tabular}
\caption{We list a selection of PDFs together with their computed 
$\kappa(Q)={\int}dx\, R_s(x,Q)$ at $Q=1.5$~GeV.
The PDFs are sorted by $\kappa(Q)$, 
and the last column shows the color key for 
Figures~\ref{fig:xs} and~\ref{fig:rscombo}.  
Note for certain PDFs $s(x)$ is linked to $\bar{u}(x){+}\bar{d}(x)$, 
so the variation of $\kappa(Q)$ is minimal. 
}  %
\label{tab:rstab}
\end{table}
Among the most robust techniques for exploring the constraints and allowed
ranged for specific PDFs in a global analysis is the Lagrange Multiplier (LM)
technique \cite{Stump:2001gu}. 
This method proceeds by constraining a PDF to
maintain a given numerical value ({\it e.g.}, a selected value for the gluon PDF
at a specific combination of $x$ and $Q$) while otherwise refitting the PDF parameters
subject to that constraint within a global fit. By continuously varying a chosen PDF
away from its central fitted value, it is then possible to quantify the corresponding
variation in $\chi^2$ for the full fit as well as individual data sets. In
Fig.~\ref{fig:Rs}, we plot the result of this procedure as explored in
Ref.~\cite{Hou:2019efy} for $R_s$ at $x=0.1$ and $Q=1.5\,\mathrm{GeV}$.
In practice, the LM technique can be used to infer a range allowed
by various fitted data sets for a specific PDF-level quantity in a QCD global
analysis. In fact, this information is plotted in Fig.~\ref{fig:Rs}, in which
points along the various curves correspond to individual refits with $R_s$ tuned
away from its respective central values within the CT18(Z) analyses in the upper and
lower panels.

\begin{figure*}[ht]
\centering
  \begin{subfigure}{a)}
        \includegraphics[width=0.95\columnwidth]{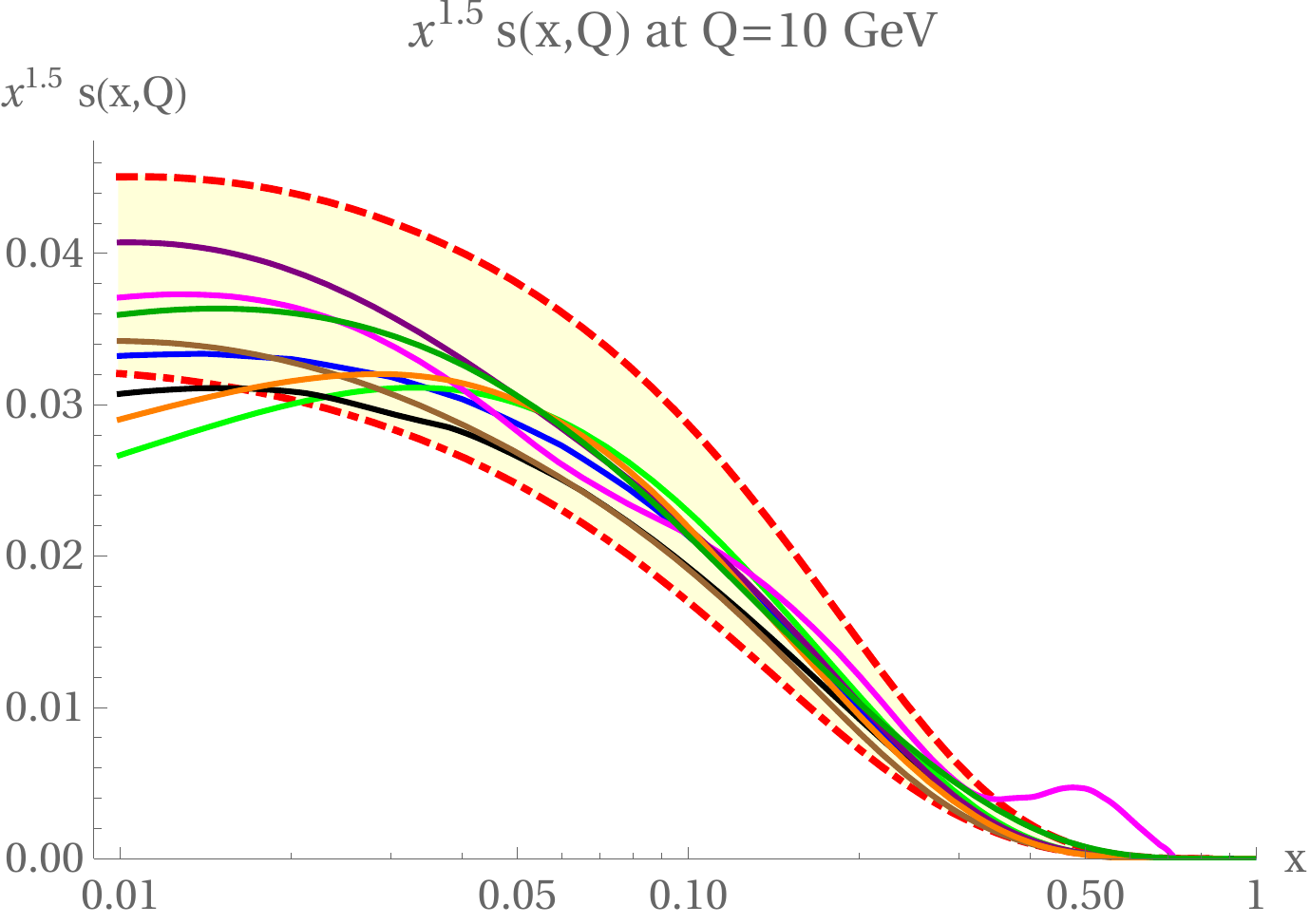}
  \end{subfigure}
  \begin{subfigure}{b)}
        \includegraphics[width=0.95\columnwidth]{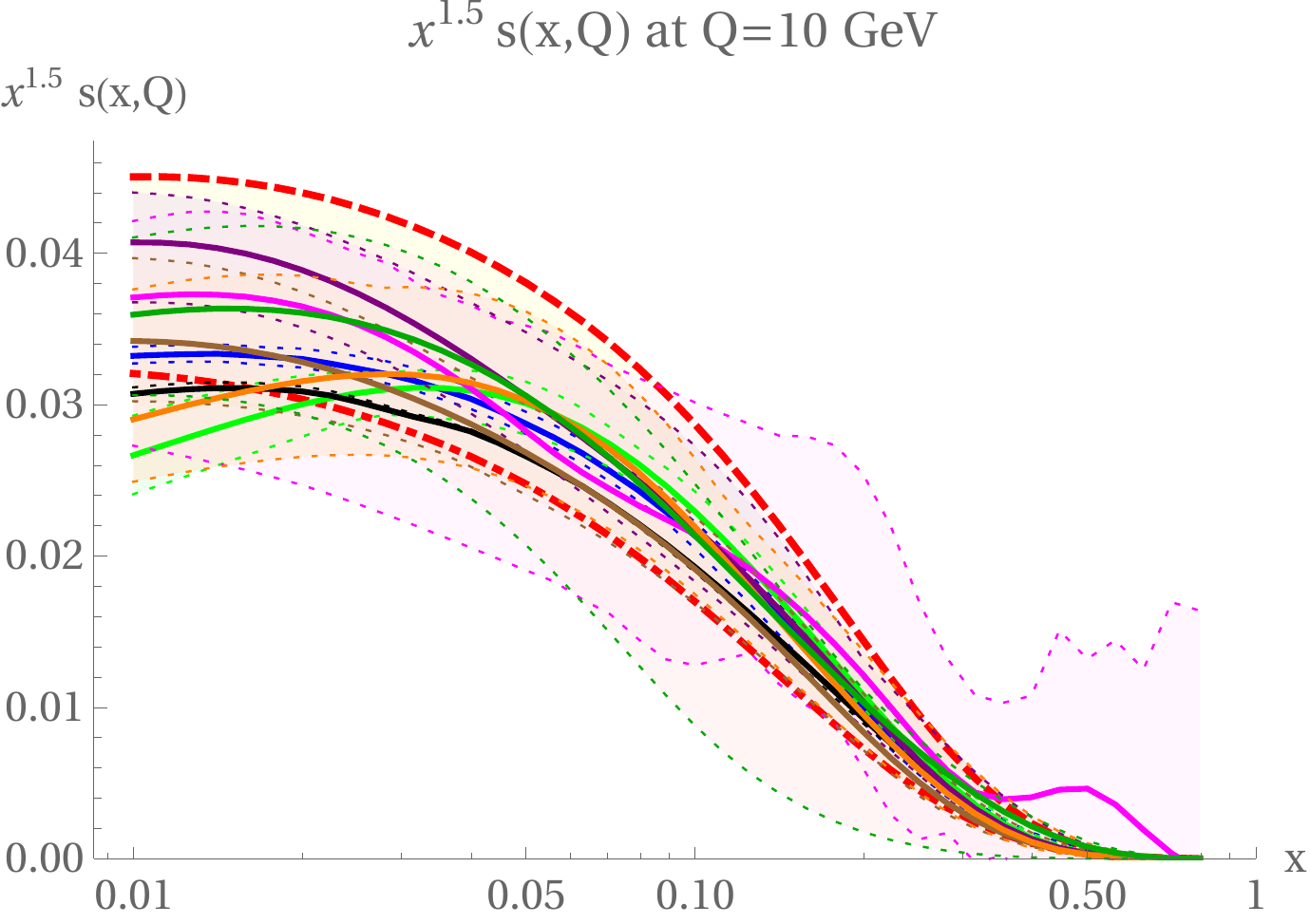}
  \end{subfigure}
\caption{a)~For a selection of PDFs, we display the 
strange quark PDF, $x^{1.5} s(x,Q)$, at $Q{=}10$~GeV.
The region between our  
\rshigh{} [enhanced strangeness] (dashed red)
and  \rslow{} [suppressed strangeness] (dot-dashed red) PDF curves is shaded
to highlight this region. 
The PDF curves are identified in Table~\ref{tab:rstab}.
\quad 
b)~Same as Fig.~a) but with the PDF uncertainties indicated
with shaded bands bounded by dotted lines. (Color online.)
}  %
\label{fig:xs}
\end{figure*}
\begin{figure*}[ht]
\centering
  \begin{subfigure}{a)}
        \includegraphics[width=0.95\columnwidth]{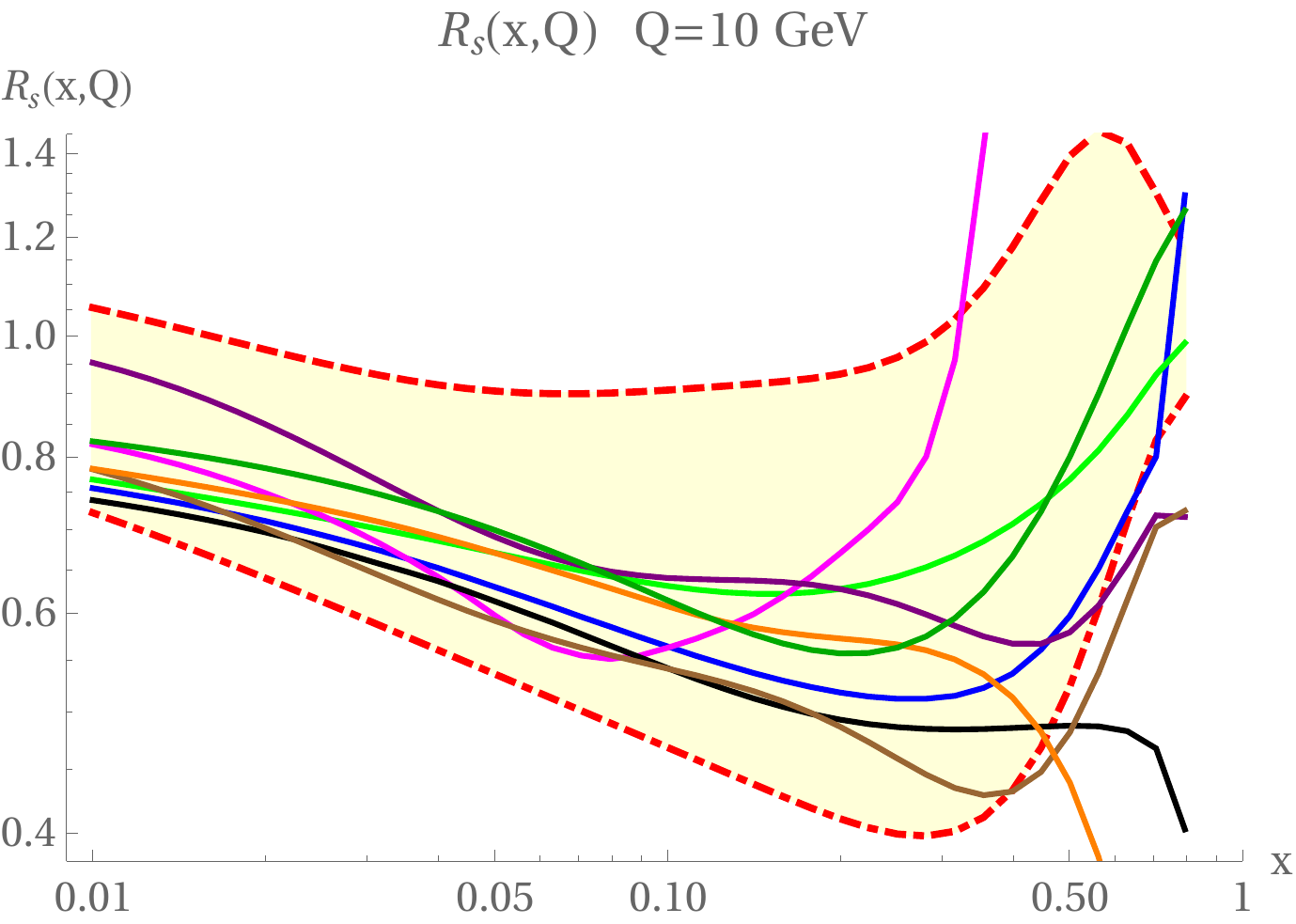}
  \end{subfigure}
  \begin{subfigure}{b)}
        \includegraphics[width=0.95\columnwidth]{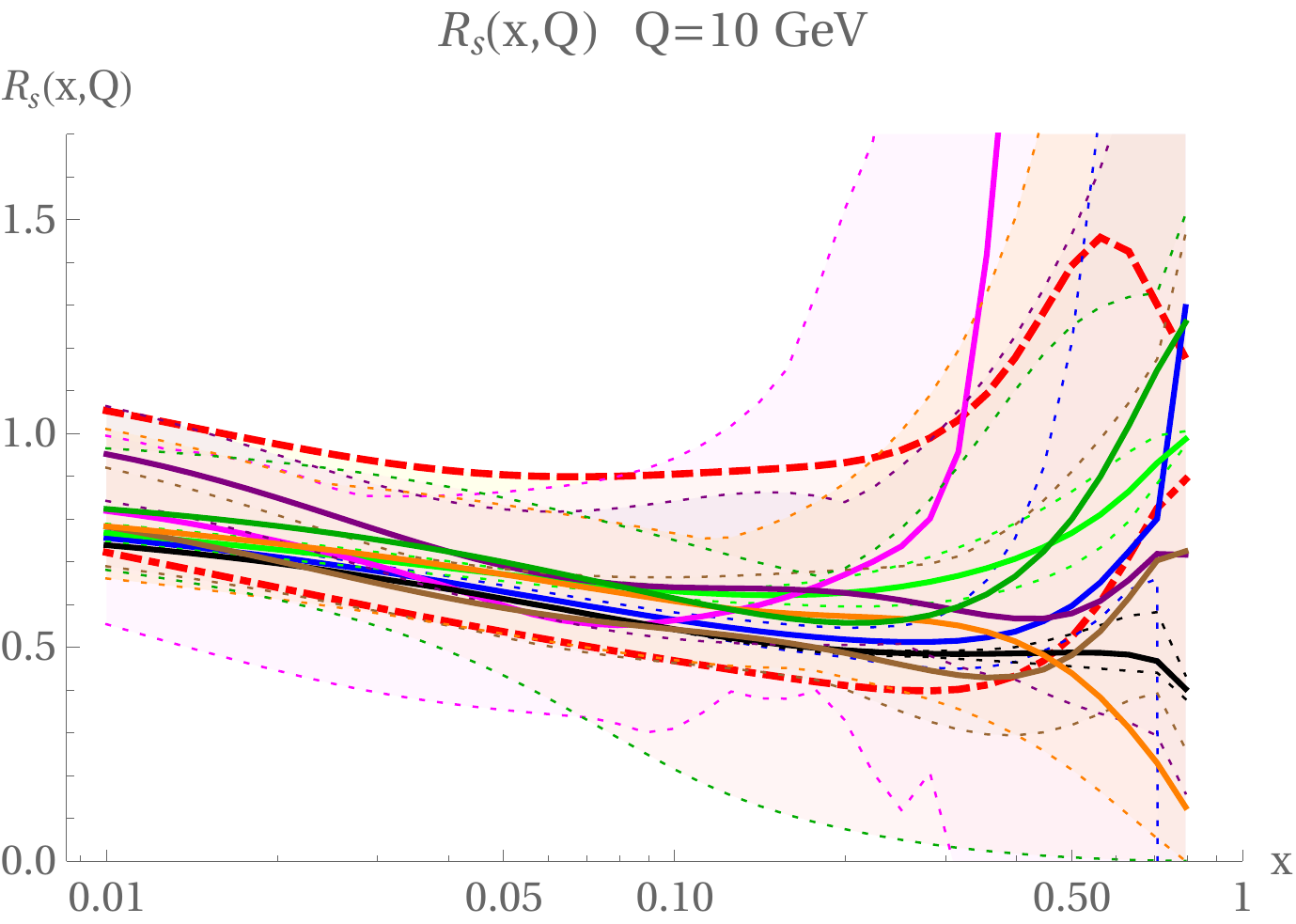}
  \end{subfigure}
\caption{
a)~For a selection of PDFs, we display the 
strange quark $R_s(x,Q)$ ratio at $Q{=}10$~GeV.
The region between our  
\rshigh{}{} [enhanced strangeness] (dashed red)
and  \rslow{} [suppressed strangeness] (dot-dashed red) PDF curves is shaded
to highlight this region. 
The PDF curves are identified in Table~\ref{tab:rstab}.
\quad
b)~Same as Fig.~a) but with the PDF uncertainties indicated
with shaded bands bounded by dotted lines. (Color online.)
}  %
\label{fig:rscombo}
\end{figure*}
\begin{figure}
    \centering
    \includegraphics[width=1.0\columnwidth]{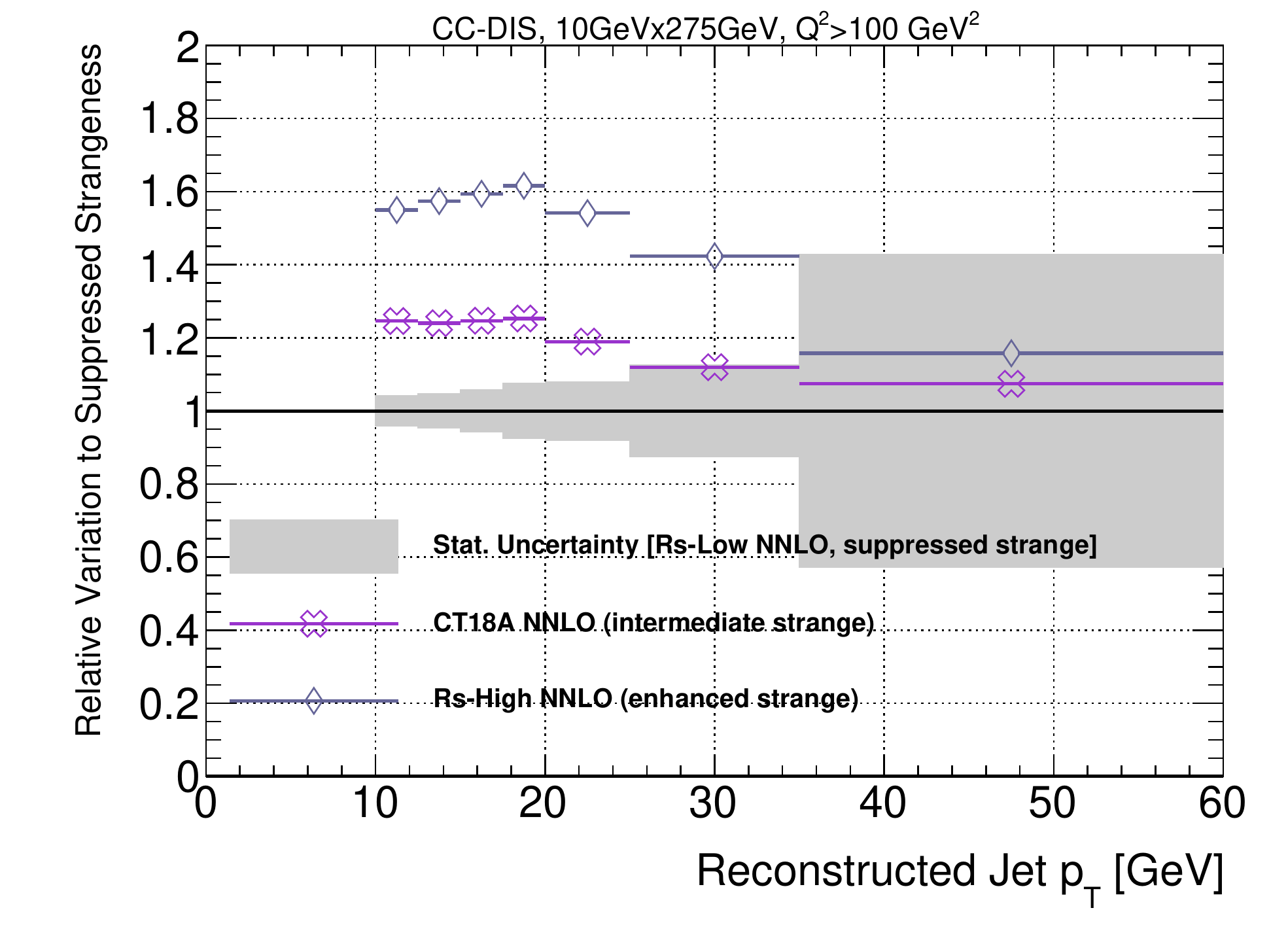}    \\
     \includegraphics[width=1.0\columnwidth]{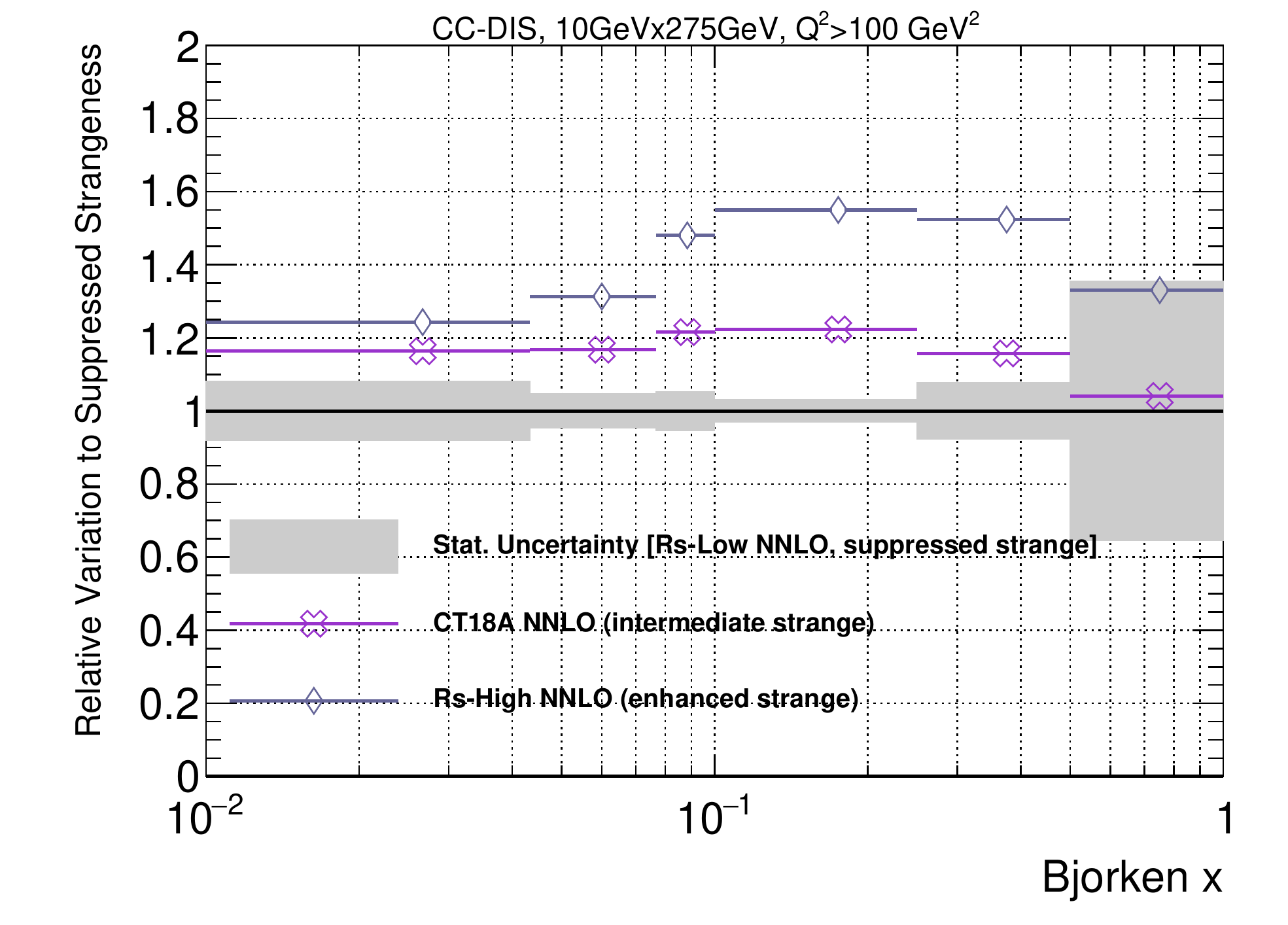}    \\
     \includegraphics[width=1.0\columnwidth]{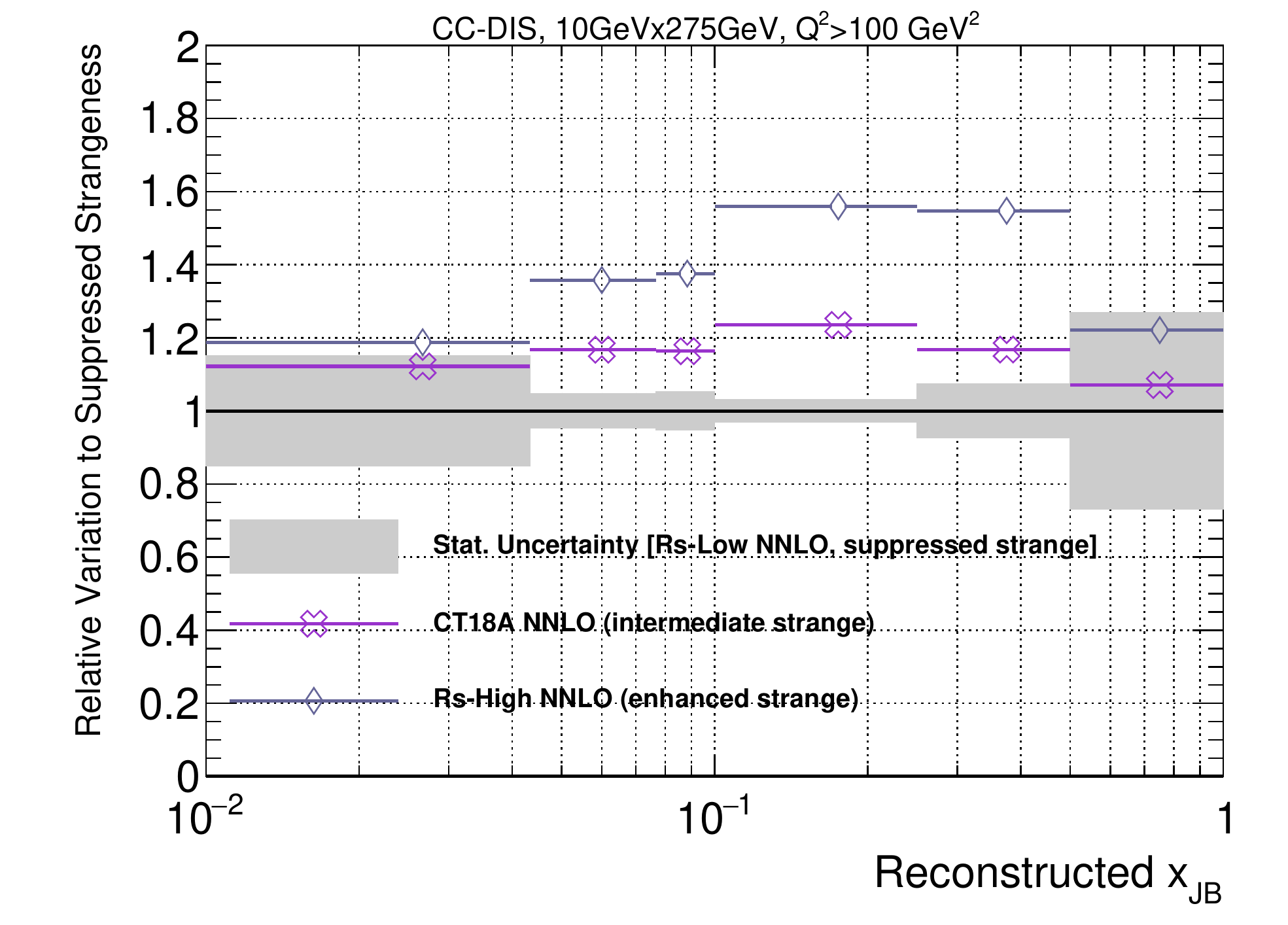}
\caption{We compare three cases: 
i)~\rslow{}{} NNLO with suppressed strangeness,
ii)~\rshigh{} NNLO with enhanced strangeness,
and 
iii)~CT18A NNLO with intermediate strangeness.
Our baseline is the \rslow{} (suppressed strangeness)
and the gray band indicates the expected statistical error on 
the reconstructed and tagged charm jet $p_T$ (top), Bjorken $x$ (middle), 
and reconstructed $x_{JB}$ (bottom) spectrum with $\mathrm{100\,fb^{-1}}$ of data. 
The points indicate the difference in expected yields   $(1+\Delta N/N)$ for  
the enhanced (\rshigh{}) and intermediate (CT18A) strangeness cases
relative to the suppressed (\rslow{}) case.
} %
\label{fig:error_comparison}
\end{figure}
\subsection{Representative PDFs}
\label{sec:pdf}

Based upon the LM scans over $R_s$ shown in  Fig.~\ref{fig:Rs}, 
we have identified  two extreme PDF sets: 
one, which we designate ``Rs-Low,'' is associated with strongly suppressed nucleon strangeness
({\it i.e.}, small $R_s\! <\! 0.5$), identified with the extreme leftmost
boundary of the upper LM scan obtained under CT18 in Fig.~\ref{fig:Rs};
and another, called ``Rs-High,'' is associated with relatively unsuppressed strange ($R_s\! \sim\! 1$),
corresponding to the rightmost boundary of the lower LM scan based on CT18Z.
These two sets allow us to delineate the acceptable range of the strange PDF. 
To provide a reference approximately midway between these two scenarios, we also
examine a PDF set with 
total strangeness intermediate between Rs-Low and Rs-High: the CT18A NNLO fit of
Ref.~\cite{Hou:2019efy}, which differs from CT18 only in including the ATLAS inclusive
$W/Z$ production data at 7 TeV.

We summarize the key properties of these three representative PDF sets below,
\begin{align}
{\mathrm{\rslow{}\  NNLO} 
\atop
{\mathrm{Suppressed} 
\atop
\mathrm{Strange} } }
&
\begin{cases}
R_s(0.1,1.5) = 0.325 \\[5pt]
\kappa(Q=1.5) = 0.37 \\
\end{cases}
\nonumber \\[5pt]
{\mathrm{CT18A\, NNLO} 
\atop
{\mathrm{Intermediate} 
\atop
\mathrm{Strange} } }
&
\begin{cases}
R_s(0.1,1.5) = 0.552 \\[5pt]
\kappa(Q=1.5) = 0.63 \\
\end{cases}
\nonumber \\[5pt]
{\mathrm{\rshigh{}\  NNLO} 
\atop
{\mathrm{Enhanced} 
\atop
\mathrm{Strange} } }
&
\begin{cases}
R_s(0.1,1.5) = 0.863 \\[5pt]
\kappa(Q=1.5) = 0.96 \\
\end{cases}
\nonumber 
\end{align}
where the arguments of  $R_s(x,Q)$ indicate 
this is evaluated for $x=0.1$ and $Q=1.5$~GeV,
and $\kappa(Q)$ is also evaluated at $Q=1.5$~GeV.

To illustrate how these three PDFs compare to commonly used PDF sets in the literature,
in Table~\ref{tab:rstab} we list selected sets along with their computed 
values for $\kappa(Q){=}{\int{}}dx\, R_s(x,Q)$.
Further comparisons are provided 
in Figure~\ref{fig:xs} which displays the strange PDF, and
in Figure~\ref{fig:rscombo} which displays the $R_s(x,Q)$.
Specifically, we present proton PDFs for 
MSTW2008~\cite{Martin:2009iq}
NNPDF31~\cite{Ball:2017nwa}
HERAPDF2.0~\cite{Abramowicz:2015mha},
CJ15~\cite{Accardi:2016qay}
and
CT18~\cite{Hou:2019efy},
and nuclear PDFs for
EPPS16~\cite{Eskola:2016oht}
and
nCTEQ15~\cite{Kovarik:2015cma}.
Table~\ref{tab:rstab} lists the LHAPDF~\cite{Buckley:2014ana} identifying information 
for the specific PDF sets used. 
There is a benefit in comparing these different PDFs. 
For example, the MSTW, NNPDF and CT proton sets are a selection of
those PDFs commonly used for precision proton analyses such as at the LHC. 
HERAPDF includes a precision analysis of the combined H1 and ZEUS DIS data, 
and CJ includes JLab data sets; both these PDFs are relevant for future EIC $ep$ DIS analysis. 
Finally, the EPPS and nCTEQ are nuclear PDFs (Pb shown), and this is important to 
consider as the EIC will use a variety of nuclear targets.

In Table~\ref{tab:rstab}
we observe that \rslow{} and \rshigh{} span the full range of $\kappa$ values,
corresponding to the significant range of $R_s$ values seen in Figs.~\ref{fig:xs}
and~\ref{fig:rscombo},
\color{black} %
and that CT18A is roughly midway between these two extremes.
However, when we include the PDF uncertainties, we see that 
the full range of $R_s$ and $\kappa$ is quite substantial and our 
\rslow{} and \rshigh{} PDF sets represent a significant, but
nonetheless conservative, range of variation. 
This point is further emphasized when we recall that other analyses of
the LHC inclusive $W/Z$ measurements have found $R_s$ values as large as
$R_s=1.13$~\cite{Aaboud:2016btc} at relatively proximal $x\! =\! 0.023$
and $Q\!\sim\! 1.38$ GeV.

In Fig.~\ref{fig:xs} we see our two extreme PDF sets also 
generally bracket the other PDFs but with some exceptions
at lower $x$ values, particularly the nuclear PDFs (nCTEQ15, EPPS16). 
In Fig.~\ref{fig:xs}-b) we also include the uncertainty bands 
for the various PDFs. Here we observe our extreme sets in fact represent
a conservative measure of the full $s(x)$ uncertainty.
Turning to Fig.~\ref{fig:rscombo}-a), we again see our two extreme PDF sets 
generally bracket central range of PDFs, 
with some exceptions at very high $x$.
In Fig.~\ref{fig:rscombo}-b) we also include the $R_s(x)$ uncertainty bands 
arising from the  PDFs uncertainty, and again  we observe our extreme sets are
a conservative measure of the full $R_s(x)$ uncertainty.
While the individual bands may be difficult to discern, 
it is clear that the total range of uncertainty is quite broad,
even compared to our \rslow{} and \rshigh{} results. 

Using Table~\ref{tab:rstab} and these two figures, 
it is possible to compare this collection of PDFs
to our three reference PDFs  \{\rslow{},CT18A,\rshigh{}\},
and then use this to estimate the relative effect 
on the resulting event-level sensitivity presented in 
Fig.~\ref{fig:error_comparison} and discussed in the following section.

\subsection{Event-level sensitivity of charm-jet production}
\label{sec:event_level_sensitivity}
For the rest of this paper, we employ only the high-impact parameter track-counting jet-tagging approach (Sec.~\ref{sec:ip3d}), to maintain charm-jet purity while sacrificing overall statistical precision. We believe this offers a reasonable, if still conservative, baseline for estimating sensitivity to intrinsic strangeness in the proton.

The EIC beam configuration studied here is expected to result in $\mathcal{O}(1000)$ events in $100\mathrm{~fb^{-1}}$ of integrated luminosity after charm tagging. We show in Fig.~\ref{fig:error_comparison} the expected precision of the tagged charm-jet spectrum. Across much of the jet \pT~region, or indeed as a function of $x_{JB}$ or true Bjorken $x$, such uncertainties would be at the level of 10\%. This is likely conservatively over-estimated, since we have demonstrated that it's possible to enhance charm tagging efficiency with modest additional effort ({\it e.g.}, single-track PID), even while we have neglected other experimental effects (triggers, knowledge of the jet energy scale, {\it etc.}) in this study.

This statistical uncertainty is to be compared to the range of variation in knowledge of the strangeness PDF for the three representative PDF scenarios displayed in Fig.~\ref{fig:error_comparison}.
Our baseline is the \rslow{} (suppressed strangeness)
and the gray band indicates the expected statistical error on 
the reconstructed and tagged charm kinematic variables 
with $\mathrm{100\,fb^{-1}}$ of data. 
Note, the statistical uncertainties are derived from the suppressed scenario, making them additionally conservative. 
We then compare with both the  CT18A (intermediate strangeness) 
and \rshigh{} (enhanced strangeness)  distributions. 
We observe that these measurements are capable of distinguishing not only 
the extreme limit of the enhanced strangeness PDF (\rshigh{}), 
but also perform well for the case of intermediate strangeness (CT18A). 
Using Table~\ref{tab:rstab} together with Figs.~\ref{fig:xs} and~\ref{fig:rscombo},
we can qualitatively estimate how PDFs with different strangeness properties 
will map on to Fig.~\ref{fig:error_comparison}.
We conclude there is strong evidence that the use of charm-tagged jets at EIC will provide new constraints on the strangeness PDF and should be part of a global analysis of strangeness within the EIC program.

The charm-jet yields described above do not include contributions from gluon-initiated processes, which is not simulated in \textsc{PYTHIA8}.
Although we neglect these backgrounds for this study, we recognize that at a future EIC detector experiment these will have to be quantified, characterized, and subtracted in order to interpret the data.

In addition to the beam energies discussed here, we have also explored lower-energy configurations ({\it e.g.}, $10 \times 100\mathrm{~GeV}$ electron-on-proton). As expected, the charm jet yields decline strongly due to decreasing production cross section combined with lower jet \pT. At such energies, the expected charm-jet yield is $\mathcal{O}(10-100)$ events. 

We have also explored kinematics that will be available at the EIC for nuclear beams whose constituents have atomic number $Z$ and mass number $A$. The per-nucleon energy of the nuclear beams is reduced by a factor of $Z/A$, which is about 0.4-0.5 for most nuclei considered at EIC. As discussed above, lower energies lead to a rapid decline in the expected statistics, so we consider the highest center-of-mass energy that can be reached for nuclear beams, which is 110 GeV per-nucleon with an 18 GeV electron beam ($\sqrt{s} \approx 90\,\mathrm{GeV}$). While the cross sections for hard processes in electron-nucleus collisions gets enhanced by a factor of $A$, the expected luminosity for nuclear beams is approximately a factor $A$ smaller, which leads to similar expected rates. The higher electron energy (18 GeV instead of 10 GeV) comes at the cost of reduced luminosity due to power limitations, which is roughly a factor of 5~\cite{EICdesign}. Thus, the expected statistical uncertainty for nuclear beams is expected to be roughly a factor of $\sqrt{5}$ larger than our nominal studies\footnote{Note that the lower hadron energy leads to a less boosted kinematics with respect to what we show in Fig.~\ref{fig:charm_jet_kinematics}. Given that the hadronic final state at mid-rapidity increases~\cite{Arratia:2019vju}, the role of the barrel hadronic calorimeter in the \met\ measurements is enhanced with respect to the higher hadron beam configuration.}, for equal running time. We thus conclude that the prospects for CC DIS charm-jet studies with nuclear beams are promising.

CC DIS measurements with nuclear beams would yield additional flavor sensitivity, for example by using deuterium or helium-3 beams, or to study nuclear effects with heavier nuclei. Given that preliminary studies of heavy-ion LHC $W/Z$ production suggest an enhanced strange component for the nuclear as well as the proton PDFs, the EIC's ability to explore a variety of nuclear beams could prove illuminating~\cite{Kusina:2016fxy,Kusina:2020lyz}. This would allow us to test the flavor dependence of anti-shadowing and EMC effects, which remains an open question (see Refs.~\cite{West:2020rlk,Arrington:2019wky,Brodsky:2019jla,Ryckebusch:2019oya,Segarra:2019gbp,Wang:2018wfz,Cloet:2012td,Schienbein:2009kk,Cloet:2009qs,Chang:2011ra,Brodsky:2004qa}). We reserve these studies for future work. 

Recent work by Borsa {\it et al}.~\cite{Borsa:2020ulb}~showed that jet production in polarized NC DIS, which they calculated to NNLO accuracy, is sensitive to quark helicity. When extended to CC DIS, those results could be compared with the precision we estimate to gauge the sensitivity to strange helicity. However, the necessity of subtracting a significant light-jet background, with its own inherent asymmetry, from the charm-tagged sample implies that care is needed to understand the ultimate reach in sensitivity of a dedicated polarization measurement. We reserve these studies for future work. 

\subsection{Possible further developments}
Our estimate for charm-jet efficiency is rather conservative, and could be improved to at least the level of modern charm-taggers in collider experiments such as the LHC that routinely yield 20--50$\%$~\cite{Aaboud:2018fhh}. We have also demonstrated the basic gains and challenges that can be expected from the use of displaced leptons and kaons from the charm-meson decays. A mature multivariate analysis would combine all the information including displaced tracks, PID, leptons, and topological ({\it e.g.}, secondary vertex) information. 

A complementary way to increment the statistical power of this channel is to lower the $Q^{2}$ cut. Given that the cross section decreases as $1/Q^{4}$, a relaxing of the selection of $>\!100$ GeV$^{2}$ to $>\!50$ GeV$^{2}$ would increment the yield substantially and gain sensitivity at lower-$x$, which further reduces the light-flavor jet background from valence quarks. The challenges associated with measuring low-$Q^{2}$ charged-current DIS are manifold, including rejection to backgrounds from photo-production and misidentified neutral-current DIS, as well as increased background from gluon-initiated processes. While most HERA studies imposed a selection on $Q^{2}>200$ GeV$^{2}$~\cite{Abramowicz:2015mha}, studies by Aschenauer {\it et al.}~\cite{Aschenauer:2013iia} showed that a lower limit of $Q^{2}>100$ GeV$^{2}$ is feasible at the EIC. Future dedicated studies should explore the limit on low $Q^{2}$, which most likely will demand highly hermetic detector systems with low thresholds. 

\section{Detector requirements}
\label{sec:detreq}

In this section, we summarize the main detector requirements to measure charm-jets in charged-current DIS: 

\begin{itemize}
    \item The reconstruction of charm jets with large radius parameter ($R=1.0$) requires tracking and calorimeter coverage extending in the positive-$z$ direction out to at least $\eta=3.5-4.0$ (Fig.~\ref{fig:charm_jet_kinematics}). A high tracking efficiency will be essential to reconstruct and tag these jets. 
    \item Given the jet kinematics (Fig.~\ref{fig:charm_jet_kinematics}) are centered around the barrel-endcap transition region of a typical collider detector, the inactive regions, material budget, and geometry have to be optimized to avoid drastically degrading the detector performance for these jets. An example of a design that achieved this is given by the ZEUS calorimeter~\cite{Derrick:1991tq}.  
    
    \item As jet production is typically in the forward direction and at lower angles to the hadron beam direction, vertex or impact parameter resolution in both the $x-y$ plane and along the $z$ direction will be essential to flavor-tagging approaches such as those described here. Significant degradation of resolution beyond the baseline of $20\mathrm{~\mu m}$ is observed to cause significant loss of charm-tagged jet yields while generally increasing light-jet efficiency. We also note that a simple optimization of tagging hyper-parameters tended to prefer track momentum thresholds down to $0.50\mathrm{~GeV}$, illustrating the need to have high efficiency for low-momentum tracks even for the purpose of selecting signal jets.

    \item We explored the use of single-track PID to enhance charm-jet tagging performance. We observed significant gains in charm-jet efficiency using baseline EIC detector PID guidance, or reasonable assumptions where such guidance was not present ({\it e.g.}, electrons and muons). Light-jet mis-tagging rates were typically better, and certainly no worse, in these approaches than in the baseline high-impact-parameter track-counting approach. This suggests that optimization and multi-variable approaches to combine information could yield strong gains in significance over the expectations reported here. It also suggests, however, that assuming worse PID efficiency and mis-identification scenarios than we employed here will reduce the value of such information in a future EIC detector. Dedicated PID system coverage and detector granularity will need to extend well into the forward region, defined above, given the size of the jets.
    \item This work implies the need for an hermetic detector, with full calorimetry coverage to reach as low a $Q^{2}$ (corresponding to low \met) as possible, while ensuring background suppression to photo-production and NC DIS. This also demands low thresholds for both tracking and calorimetry, as well as calorimetry resolution (the tracking resolution is subdominant). One example of the importance of hermeticity and its impact on jet resolution will be the trigger; while we did not explore trigger algorithms in this work, an \met-based CC DIS trigger algorithm will benefit from strong coverage and finer granularity, as any trigger decisions will necessarily use lower resolution than is available in a fully-calibrated offline environment. The efficiency of the \met\ requirement (Sec.~\ref{sec:event_selection}) is sufficiently below unity that degrading this resolution further has strong implications for potential trigger efficiency. 
\end{itemize}

\section{Summary and conclusion}
\label{sec:conclusions}

We have explored the experimental feasibility of charm-jet cross-section measurements in charged-current DIS at the future Electron-Ion Collider. We use parametrized detector simulations with the \textsc{Delphes} package with baseline parameters for the EIC detectors. We estimated the performance of an high-impact-parameter track-counting algorithm to tag charm jets. We also explored the potential of particle identification to increment tagging efficiency. Our feasibility studies suggest that the prospects for constraining unpolarized nucleon strangeness are rather promising in this channel. These goals represent a challenge that demands high luminosity as well as a well-designed EIC detector with good capabilities for measuring displaced vertices, particle ID, jets, and missing-transverse energy. As such, it represents a robust platform on which to inform the design of the EIC detectors. 

The charm-tagging performance studies advanced in this work have the potential to extend the rapidly emerging field of jet studies for the future EIC~\cite{Arratia:2020nxw,Arratia:2020ssx,Borsa:2020ulb,Peccini:2020tpj,Guzey:2020gkk,Guzey:2020zza,Kang:2020xyq,Li:2020sru,Gutierrez-Reyes:2019vbx,Arratia:2019vju,Page:2019gbf,Gutierrez-Reyes:2019msa,Zhang:2019toi,Aschenauer:2019uex,Hatta:2019ixj,Mantysaari:2019csc,DAlesio:2019qpk,Kishore:2019fzb,Kang:2019bpl,Roy:2019hwr,Salazar:2019ncp,Gutierrez-Reyes:2018qez,Boughezal:2018azh,Klasen:2018gtb,Dumitru:2018kuw,Liu:2018trl,Zheng:2018ssm,Sievert:2018imd,Klasen:2017kwb,Hinderer:2017ntk,Chu:2017mnm,Abelof:2016pby,Hatta:2016dxp,Dumitru:2016jku,Boer:2016fqd,Dumitru:2015gaa,Hinderer:2015hra,Altinoluk:2015dpi,Kang:2013nha,Pisano:2013cya,Kang:2012zr,Kang:2011jw,Boer:2010zf}. In particular, charm-jet tagging approaches could be applied to neutral current boson-gluon fusion ({\it e.g.}, see pp.~289 of Ref.~\cite{Aidala:2020mzt}) or photo-production processes. Exploiting CC DIS charm-jet measurements to constrain the nucleon's quark-gluon structure will also require continued advances in precision QCD and global analyses in order to ensure the stability of
the eventual PDF extractions we envision.

\section*{Code availability}
The Delphes configuration file for the EIC general-purpose detector used in this work can be found in: 
\url{https://github.com/miguelignacio/delphes_EIC}{}

\section*{Acknowledgements} \label{sec:acknowledgements}
We are grateful to Pavel Nadolsky for helpful inputs, especially related to PDF choices and theory implications. We thank Oleg Tsai for insightful discussions
on calorimetry technology for EIC detectors. We thank Daniel de Florian for commenting our manuscript. We thank the members of the EIC User Group for many insightful discussions during the Yellow Report activities. 
M.A and Y. F acknowledge support through DOE Contract No. DE-AC05-06OR23177 under which JSA operates the Thomas Jefferson National Accelerator Facility. 

T.J.H., S.J.S., and F.O.\   acknowledge  support through US DOE grant DE-SC0010129. 
T.J.H.~also acknowledges support from a JLab EIC Center Fellowship.
We gratefully acknowledge SMU's Center for Research Computation for their support and for the use of the SMU ManeFrame~II high-performance computing cluster, which enabled a portion of the simulation and analysis work in this paper.

\FloatBarrier

\bibliographystyle{apsrev4-1} 

\bibliography{bibliography.bib,extra.bib}

\end{document}